\documentclass[aps,prd,superscriptaddress,nofootinbib,amsmath,amsfonts,groupedaddress]{revtex4-2}

\usepackage{setspace}
\usepackage{graphicx}
\usepackage{subfig}
\usepackage{epsfig}
\usepackage{bm}
\usepackage{amssymb}
\usepackage{float}
\usepackage{amsmath}
\usepackage{dcolumn}
\usepackage[colorlinks]{hyperref}
\usepackage[usenames,dvipsnames]{color}
\usepackage{enumitem}

\hypersetup{ breaklinks=true, pdfstartview={FitH}, colorlinks=true, linkcolor=blue, citecolor=red, filecolor=magenta, urlcolor=blue, anchorcolor=green, linktocpage=true }

\def\doi{http://doi.org}

\newcommand{\be}{\begin{equation}}
\newcommand{\ee}{\end{equation}}
\newcommand{\beano}{\begin{eqnarray*}}
\newcommand{\eeano}{\end{eqnarray*}}
\newcommand{\ba}{\begin{eqnarray}}
\newcommand{\ea}{\end{eqnarray}}

\begin{document}

\title{ Dynamical analysis of a hyperbolic solution in Scale-covariant theory}
\author{Shaily}
\email{shailytyagi.iitkgp@gmail.com}  
\affiliation{Department of Mathematics, Netaji Subhas University of Technology, New Delhi-110078, India}
\author{J. K. Singh}
\email{jksingh@nsut.ac.in}
\affiliation{Department of Mathematics, Netaji Subhas University of Technology, New Delhi-110078, India}
\author{Joao R. L. Santos}
\email{joaorafael@df.ufcg.edu.br}
\affiliation{UFCG-Universidade Federal de Campina Grande-Unidade Academica de Fisica, 58429-900 Campina Grande, PB, Brazil}
\author{M. Zeyauddin}
\email{mdzeyauddin@gmail.com}  
\affiliation{Department of General Studies (Mathematics), Jubail Industrial College, Jubail Industrial City, 31961, Kingdom of Saudi Arabia }
\begin{abstract}
\begin{singlespace}

\noindent
We study an isotropic flat FLRW-model in Scale-covariant theory of gravity $ f_{\gamma \delta}(\phi) $ \cite{Canuto:1977zz} which is explained in terms of ordinary and covariant differentiation of scalar field $ \phi $. As we know the deceleration parameter is time-dependent, so we consider the deceleration parameter $ q $ as the function of $ t $. Using this methodology, we find all the important cosmological factors in terms of a hyperbolic function of the cosmic time $ t $. In turn, we create the model having the behavior of the late-time universe, which is ever accelerated expanding and faces a Big Freeze at the end.  The model shows the quintessence dark energy model from early to late times. We compute the constrained values of Hubble parameter $ H_0=70.979^{+0.021}_{-0.0043} $ and the model parameter $ n=1.24079^{+0.00015}_{-0.00079} $ using joint analysis of the $ OHD $ data of 77-points and Pantheon bin data. The model exhibits point-type singularity, beginning with a point of zero volume, infinite energy density, and temperature. Furthermore, we obtain the present deceleration parameter $ (q_0) \approx {-0.54} $. Also, we examine the ultimate behavior of our model by properly analyzing energy conditions, cosmographical parameters, and Statefinder diagnostic. Finally, the proposed model behaves like a quintessence dark energy model.

\end{singlespace}

\end{abstract}

\maketitle

\section{Introduction} 

\qquad Problems like dark matter or missing matter in cosmology can not be resolved by General Relativity (GR). So the scalar-tensor theories have a greater role to play in describing many aspects of gravitational physics and cosmology, which are obtained by some modifications in GR. A convenient set of representations is provided by these scalar-tensor theories for the observational limits. One of the widely accepted theories is the Scale-Covariant theory, which was initiated by Canuto et al. \cite{Canuto:1977zz, Canuto:1977dp} to measure space-time distances by using scale transformation. In this theory, they have used the physics of different dynamical systems and have formulated this scale-covariant theory. Another arbitrary gauge function is determined by gauge condition which corresponds to each dynamical system of units. In scale transformation, the field equations of generalized Einstein are invariant. The gravitational constant $ G $ \cite{Wes, Will:1984qgz} to be time-dependent is indicated by this theory. Interestingly, the field equations and other physical parameters are measured in gravitational units as well as in atomic units. Here, we consider the space-time for Einstein's units at $ d\bar{s}^{2}=\bar{g_{ij}}dx^{i}dx^{j} $ whereas the line element for other units is given by the relation $ ds=\phi^{-1}(x)d\bar{s} $, here indices i and j take values 1,2,3,4. Also we use the transformation between the metric tensors $ \bar{g_{ij}} $ and $ g_{ij} $ and it can be given by $ \bar{g_{ij}}=\phi^{2}g_{ij} $. Here the metric $ \bar{g_{ij}} $ is macroscopic and the metric $ g_{ij} $ is microscopic. The variable gauge function $ \phi $ is considered here.

Singh et al. \cite{Singh:2013gwj} have obtained a solution in scale covariant theory for the Bianchi II line element by taking the EoS parameter as time-dependent as well as the constant deceleration parameter. In this way, a dark energy cosmological model is obtained. Sharma et al. \cite{Sharma:2014kzy} have found a string cosmological model in scale covariant theory for the Bianchi II line element. They have considered a magnetic field in the model.  A big-bang model of the cosmos is represented by the model. Reddy et al. \cite{Reddy:2007zzf}, have obtained some models using scale covariant theory. They have explored several features of the model. In this theory, Zeyauddin et al. \cite{Zeyauddin:2012bg}, obtained a model for Bianchi VI line element and discovered an exact solution under some specified assumptions. Several authors have done a lot of work in the modified theories of gravity \cite{Beesham:1998ih, Zeyauddin:2010zz, Tawfik:2019dda, Chattopadhyay:2020mqj, Tawfik:2017ngn}. Tawfik et al. have studied the quark-hadron phase transition and equation of state with the effect of the bulk viscosity \cite{Tawfik:2011sh, Tawfik:2010bm} and also discussed thermodynamical behavior of the universe for different eras \cite{Tawfik:2010ht, Tawfik:2019jsa, Tawfik:2011gh}. 

Pradhan et al. \cite{Pradhan:2006rn}, used the variable deceleration parameter (DP) to establish a solution to the Einstein field equation for LRS Bianchi I. Pradhan \cite{Pradhan:2012he} have used the concept of variable DP to produce a time-dependent scale factor in their research. They were able to create models of the cosmos that show a phase shift from the early decelerating to the current accelerating phase using this hypothesis. According to Akarsu et al., \cite{Akarsu:2011zd}, the deceleration parameter should vary linearly with time. Many exact solutions achieved with a constant deceleration parameter are generalized by their law. In the presence of an anisotropic fluid, Pradhan \cite{Pradhan:2012zt} constructed two accelerating DE models in Bianchi VI. He used a specific type of scale factor that is time-dependent for the deterministic solution. Here DP is also time-dependent. The Deceleration Parameter $ q $ transits from deceleration to acceleration. Nagpal et al. \cite{Nagpal:2019vre} have transformed the Hubble Parameter $ H $ and the Deceleration Parameter $ q $ as a function of $ z $ to get a hyperbolic solution in $ f(R,T) $ gravity. Recently many authors have done a lot of work on modified theories of gravity \cite{Singh:2018xjv, Singh:2022jue, Singh:2023gxd, Singh:2024kez}.

In this paper, we assemble our work in the following manner. Sect. I have been furnished with a brief formal introduction to the Scale-Covariant theory of gravity and the scheme for cosmic acceleration at an infinite time which has been validated by numerous investigations. In Sect. II, we look at the Einstein field equations and use a hyperbolic parametrized technique to find solutions. By operating the three different observational datasets $ OHD $ of 77 data points, $ Pantheon $ bin data of 48 points, and joint datasets $ H(z)+ Pantheon $, the model parameters are constrained for the comprehensive examination of the performance of physical parameters in Sect. III. In Sect. IV, the physical features of the parametric model have been discussed concisely by applying statistical data analysis. Moreover, the viability of the model has been illustrated through energy conditions. In Sect. V, we perform different diagnostic techniques to authenticate our model as a dark energy model. In sect. VI, we validate the second law of thermodynamics, and finally, the consistency of the model has been evaluated utilizing data analysis using certain statistics and conclude the consequences of the derived model.

\section{Basic Equations and its dynamics}
\qquad Canuto et al. \cite{Canuto:1977zz} draw up the scale-covariant theory of gravitation which is an effective alternative to general relativity. Within the framework of scale covariant theory, Einstein’s field equations are viable in gravitational units through physical quantities counted in atomic units. For two systems of units, we have a conformal transformation for metric tensors 
\begin{equation}\label{0}
\bar{g}_{\gamma \delta}= \phi^2 g_{\gamma \delta},
\end{equation}
where indices $ \gamma $, $ \delta $ take their values $ 1,2,3,4 $, the bar indicates gravitational units, and unbar is used for atomic quantities. The scalar function $ \phi $ satisfying $ 0 < \phi < \infty $  in its most general formulation may be stated as a function of all space-time coordinates. Now, including a matter Lagrangian, we state the action principle as follows: 
\begin{equation} \label{1}
I=\int (- \phi^2 R+c \phi^4 + 16 \pi G L_m) \sqrt{-g}  d^{4}x,
\end{equation} 
where $ c $ is set to $ -\frac{2 \Lambda}{\phi^2} $ and $ L_m $ is matter Lagrangian. Further, the field equations in Scale Covariant theory can be written as \cite{Canuto:1977zz, Canuto:1977dp}
\begin{equation}\label{2}
R_{\gamma \delta}-\frac{1}{2} R g _{\gamma \delta}+f_{\gamma \delta}(\phi) =8\pi G T_{\gamma \delta}+\Lambda (\phi)g_{\gamma \delta},
\end{equation}
where
\begin{equation}\label{3}
f_{\gamma \delta}(\phi)=\frac{1}{\phi^{2}}\Big[ 2\phi \phi_{\gamma;\delta}-4\phi_{\gamma}\phi_{\delta}-g_{\gamma \delta}(2\phi\phi^{\lambda}_{;\lambda}-\phi^{\lambda}\phi_{\lambda})\Big],
\end{equation}

Here $R_{\gamma \delta}$, $R$, $G$, $T_{\gamma \delta}$ and $\Lambda$ stand for Ricci tensor, Ricci curvature, the gravitational constant, energy-momentum tensor cosmological constant respectively. The term $ \phi_{\gamma}$ is an ordinary derivative, while covariant differentiation is denoted by a semi-colon. In literature, there are modified theories that appreciate minimal as well as non-minimal coupling between curvature or matter invariants but we have considered the scale-covariant theory which involves a non-minimal coupling between a gauge function and Ricci scalar \cite{Ellis:2013iea}. Without this coupling, models are out of observation constraints for the scalar factor. The actual physical motivation to consider this theory is to establish a dark energy model which provides more compatible results and compare our results with the recent observational Planck's data. 

For a flat FLRW space-time  
\begin{equation}\label{4}
ds^2=-a^{2}(t)(dx^2+dy^2+dz^2)+dt^2.
\end{equation}
Where $ a $ is the scale factor which depends on time. Also, the energy-momentum tensor for perfect fluid can be written as
\begin{equation}\label{5}
T_{\gamma \delta} = \left(\rho +p \right)u_{\gamma}u_{\delta} - p g_{\gamma \delta},
\end{equation}
where $ \rho $, $ p $, and $ u^{\gamma} $ represent the energy density, pressure, and the four-velocity vector respectively. In co-moving coordinate system, $ u^{\gamma}u_{\gamma}=1 $ and $u^{\gamma}u_{\delta}=0$. The Einstein field equation for a flat FLRW space-time with zero cosmological constant $\Lambda$ can be found by expanding the tensor Eqs. (\ref{2}, \ref{3}) as 
\begin{eqnarray}\label{6}
2\dot{H}+3H^2+6H\frac{\dot{\phi}}{\phi}+2\frac{\ddot{\phi}}{\phi}-\frac{\dot{\phi}^2}{\phi^2}=-8\pi G p,
\end{eqnarray}
\begin{eqnarray}\label{7}
3H^2+6H\frac{\dot{\phi}}{\phi}+3\frac{\dot{\phi}^2}{\phi^2}=8\pi G \rho,
\end{eqnarray}
Here we take $ 8\pi G =1 $. The continuity equation which is a consequence of the field Eqs. (\ref{6}) and (\ref{7}), in the scale covariant theory is given by  \cite{Canuto:1977dp, beesham, Singh:2019fga}
\begin{equation}\label{7a}
 \dot{\rho}+(\rho+p)u^{\gamma}_{;\gamma}+\rho\Big(\frac{\dot{G}}{G}+\frac{\dot{\phi}}{\phi}\Big)+3p\frac{\dot{\phi}}{\phi}=0,
\end{equation}
which leads
\begin{equation}
    \dot{\rho}+3H(\rho+p)=-\rho\Big(\frac{\dot{G}}{G}+\frac{\dot{\phi}}{\phi}\Big)-3p\frac{\dot{\phi}}{\phi}.
\end{equation}
The presence of the scalar field affects the equation of continuity of the matter field which is clarified in the appendix. As it is observed that today's universe is in a state of accelerating expansion which is verified by the recent observations of $ SNeIa $ and $ CMB $ anisotropies \cite{VargasdosSantos:2015kfv,SupernovaSearchTeam:1998fmf,SupernovaSearchTeam:2001qse,SupernovaCosmologyProject:1997zqe,SupernovaCosmologyProject:1998vns,SupernovaCosmologyProject:2003dcn,SupernovaSearchTeam:2003cyd,HighZSNSearch:2005xhg,WMAP:2003ivt,Boomerang:2000efg,Hanany:2000qf,SupernovaSearchTeam:2004lze,Riess:2006fw,SNLS:2005qlf,Davis:2007na,Amendola:2002kd,Blandford:2004ah,Tawfik:2019qyd,Tawfik:2021rvv}. According to these observations, the deceleration parameter $ q $ must be time-dependent. To reduce the mathematical complexity and find the solution of the model which is accelerated expanding at present, we assume the deceleration parameter $ q $ as a function of $ t $ as:
\begin{equation}\label{8}
q=-\frac{a\ddot{a}}{\dot{a}^{2}}=-\left(\frac{\dot{H}+H^2}{H^2}\right)=\kappa(t).
\end{equation}
The above Eq. (\ref{8}) can be transformed to
\begin{equation}\label{9}
\frac{\ddot{a}}{a}+\kappa \frac{\dot{a}^2}{a^2}=0.
\end{equation}
Here $ a $ is the scale factor and is time-dependent. The following relationship can also be assumed as 
\begin{equation}\label{10}
\kappa=\kappa(t)=\kappa(a(t)).
\end{equation}
The above assumption of $ \kappa $ is possible only if there is a one-one correspondence between $ t $ and $ a $. This idea is valid as both $ t $ and $ a $ are increasing functions. Using Eqs. (\ref{8}) and (\ref{9}), we get 
 \begin{equation}\label{11}
\int{e^{\int{\frac{\kappa}{a}da}}da}=t+\kappa_0,
\end{equation}
where $ \kappa_0 $ is a constant for integrating. We choose $\int{\frac{\kappa}{a}da} $ in the following way without losing generality, such that the preceding Eq. (\ref{11}) can be integrated as follows:
\begin{equation}\label{12}
\int{\frac{\kappa}{a}da}=\ln{f(a)}.
\end{equation}
Without sacrificing the generality of the results, we consider the above equation. As a result of Eqs. (\ref{11}) and (\ref{12}), we get
\begin{equation}\label{13}
\int{f(a)da}= t+\kappa_0.
\end{equation}
For the physically valid solution, we assume $ f(a) $ as
\begin{equation}\label{14}
f(a)=\frac{na^{n-1}}{\beta\sqrt{1+a^{2n}}},
\end{equation}
where $ \beta $ and $ n $ are arbitrary positive constants. Using Eq. (\ref{14}) and integrating Eq. (\ref{13}) \textit{w.r.t.} the scale factor $ a $ yields
\begin{equation}\label{15}
a(t)= \sinh^{\frac{1}{n}}(\beta t).
\end{equation}
The above approach for a time-dependent deceleration parameter is used in many studies and this Ansatz is consistent with the recent observational datasets \cite{Chawla:2012it,Mishra:2013lja,Ahmed:2013bdq,Tiwari:2017emt}. Using Eq. (\ref{15}), we calculate $ H(t) $ and $ q(t) $ as follows: 
\begin{equation}\label{16}
H(t)= \frac{\beta \coth(\beta t)}{n},
\end{equation}
\begin{equation}\label{17}
q(t)=-\frac{a\ddot{a}}{\dot{a}^{2}}= n\left[1-\tanh^2(\beta t)\right]-1.
\end{equation}

The deceleration parameter is dependent on time $ t $ in Eq. (\ref{17}), and universe inflation is dependent on the sign of $ q $. The positive and negative signs of $ q $ explain the Universe's deceleration and acceleration in phases, respectively. The model decelerates and accelerates according to $ t>\frac{\tanh^{-1}(1-\frac{1}{n})^{\frac{1}{2}}}{\beta}  $ and  $ t<\frac{\tanh^{-1}(1-\frac{1}{n})^{\frac{1}{2}}}{\beta}  $  respectively for the $a(t)$ parametrization (\ref{15}). The model has a transition in phase when $ q=0 $ for $ t=\frac{1}{\beta}\,\tanh^{-1}(1-\frac{1}{n})^{\frac{1}{2}} $. The fact that the cosmos accelerates in late time, resulting in a steady expansion in the past, is self-evident \cite{SupernovaSearchTeam:1998fmf, SupernovaCosmologyProject:1998vns}, therefore the parametrization of the average scale factor makes sense. Now applying the relation 
\begin{equation}\label{18}
 \frac{a}{a_0}=\frac{1}{1+z},
\end{equation}
where $ a_0 $ is the current value of $ a $. The parameters $ t(z) $, $ H(z) $ and $ q(z) $ can be evaluated as functions of redshift $ z $ as  
\begin{equation}\label{19}
t(z)=\frac{\sinh^{-1}\sqrt{\frac{n-(1+q_0)}{(z+1)^{2n}(q_0+1)}}}{\beta}.
\end{equation}
\begin{equation}\label{20}
H(z)=\frac{\beta\,\coth\left(\sinh^{-1}\sqrt{\frac{n-(q_0+1)}{(z+1)^{2n}(q_0+1)}}\right)}{n},
\end{equation}
\begin{equation}\label{21}
q(z)=n-1-n\Bigg[\tanh\Bigg(\sinh^{-1}\sqrt{\frac{n-(1+q_0)}{(z+1)^{2n}(q_0+1)}} \Bigg)\Bigg]^2.
\end{equation}
Simplifying Eq. (\ref{21}), we get 
\begin{equation}\label{22}
q(z) = -1 + n +\frac{(1 + q_0 - n) n}{\left[ -1 - q_0 + (1 + q_0) (1 + z)^{2 n} + n\right] },
\end{equation} 
where $ q_0 $  is the current value of $ q $.

Now the gauge function $ \phi $ can be taken as directly proportional to the scale factor $ a $ as \cite{Canuto:1977zz, Canuto:1977dp, Wes}  
\begin{eqnarray}\label{23}
\phi=\phi_0 a^{\alpha}=\phi_0 \left[\sinh(\beta t)\right]^{\alpha/n},
\end{eqnarray}
where the scale covariant parameter $ \alpha $ and $\phi_0$ both are arbitrary constants.

The Eqs. (\ref{6}) and (\ref{7}) yield the expressions for $ \rho $, $ p $ and $ \omega $ as:
\begin{eqnarray}\label{24}
\rho=\frac{3 (\alpha +1)^2 \beta ^2 ((q_0+1) ((z+1)^{2 n}-1)+n)}{8 \pi  G n^2 (n-q_0-1)}, 
\end{eqnarray}

\begin{eqnarray}\label{25}
p=\frac{\beta ^2 ((\alpha  (\alpha +6)+3) (-n+q_0+1)+(q_0+1) (-\alpha  (\alpha +6)+2 (\alpha +1) n-3) (z+1)^{2 n})}{8 \pi  G n^2 (n-q_0-1)},
\end{eqnarray}
and EoS parameter can be calculated as
\begin{eqnarray}\label{26}
\omega=\frac{(\alpha  (\alpha +6)+3) (-n+q_0+1)+(q_0+1) (-\alpha  (\alpha +6)+2 (\alpha +1) n-3) (z+1)^{2 n}}{3 (\alpha +1)^2 ((q_0+1) ((z+1)^{2 n}-1)+n)}.
\end{eqnarray}
Now using the observational data $ OHD $, $ Pantheon $ bin dataset, and their joint data we find the best-fit values of the model parameter.

\section{Statistical analysis of the scale covariant parameter} \label{sec:floats} 

\qquad The most recent observational cosmology could be crucial in understanding early evolution, reheating after inflation, primordial nucleosynthesis, the development of structures, as well as the many features of dark matter and energy of the universe by implementing ray detectors and cosmic mechanisms in modern cosmology. Other observational datasets derive from the map of galaxy distribution, and it encodes present fluctuations in the Universe, such as $ SDSS $. We also have different other datasets in this context. The examples can be followed. The Big Bang hypothesis is supported by CMBR. The term QUASARS refers to the interaction between observers and quasars. $ BAO $ studies large-scale structures of the Universe to better comprehend Dark Energy. $ SNeIa $ is a more appreciating technique to calculate cosmic distances, which is named standard candles.

We make a comparison of our model with $ \Lambda $CDM in this study by using error bar graphs of OHD and $\mu(z)$ from the Hubble dataset of 77 data points \cite{Singh:2022nfm, Singh:2023ryd}, $Pantheon$ bin dataset \cite{Pan-STARRS1:2017jku, Riess:1998dv, Jha:2005jg, Hicken:2009df, Contreras:2009nt, SDSS:2014irn} and their joint datasets. Using the statistical analysis method, the model parameter $ n $ and $ H_0 $ (the current Hubble parameter) in the model were restricted by using bin data (48 points), recent observational datasets $ OHD $ of 77 data points, and joint datasets $ OHD+ Pantheon $. The physical properties of our model were described using this model parameter $ n $ restricted values. The error bar graphs of Pantheon bin data, the Hubble dataset of 77 data points \cite{Shaily:2024nmy, Balhara:2023mgj, Balhara:2023owb, Singh:2022ptu, Shaily:2024xho}, and joint datasets $ OHD+ Pantheon $ in Fig. \ref{Fig:1}, indicate that all the panels are well fitted, especially at early stages in the Universe's development, when connecting our model with observations. In Fig. \ref{Fig:1}(a) the approximate present value of Hubble parameter for $ OHD $, $ Pantheon $ and $ H(z)+ Pantheon $ are $ 72.114 $, $ 72.191 $ and $ 72.113 $ respectively.

The likelihood contours are plotted using the MCMC method with the emcee module in Python at the confidence levels of $ 68.27\% $ and $ 95.45\% $ with the errors $ 1\sigma $ and $ 2\sigma $ in the $ n $-$ H_ 0 $ plane. These plots are based on good fitting points and fit the model with the recent observational datasets $ OHD $ of 77 data points, $ Pantheon $ bin data, and joint datasets $ OHD+ Pantheon $, respectively.

$ OHD $ (the Hubble data) is a functional and straightforward instrument for assessing cosmological ideas in a stream of current discoveries because it is intimately linked to the universe's expanding history. The Hubble parameter data also aid us in highlighting the importance of the universe's dark feature, which covers issues like dark matter, dark energy, and dark ages. The Hubble parameter $ H $ is given as: 

\begin{figure}\centering
	\subfloat[]{\label{a}\includegraphics[scale=0.40]{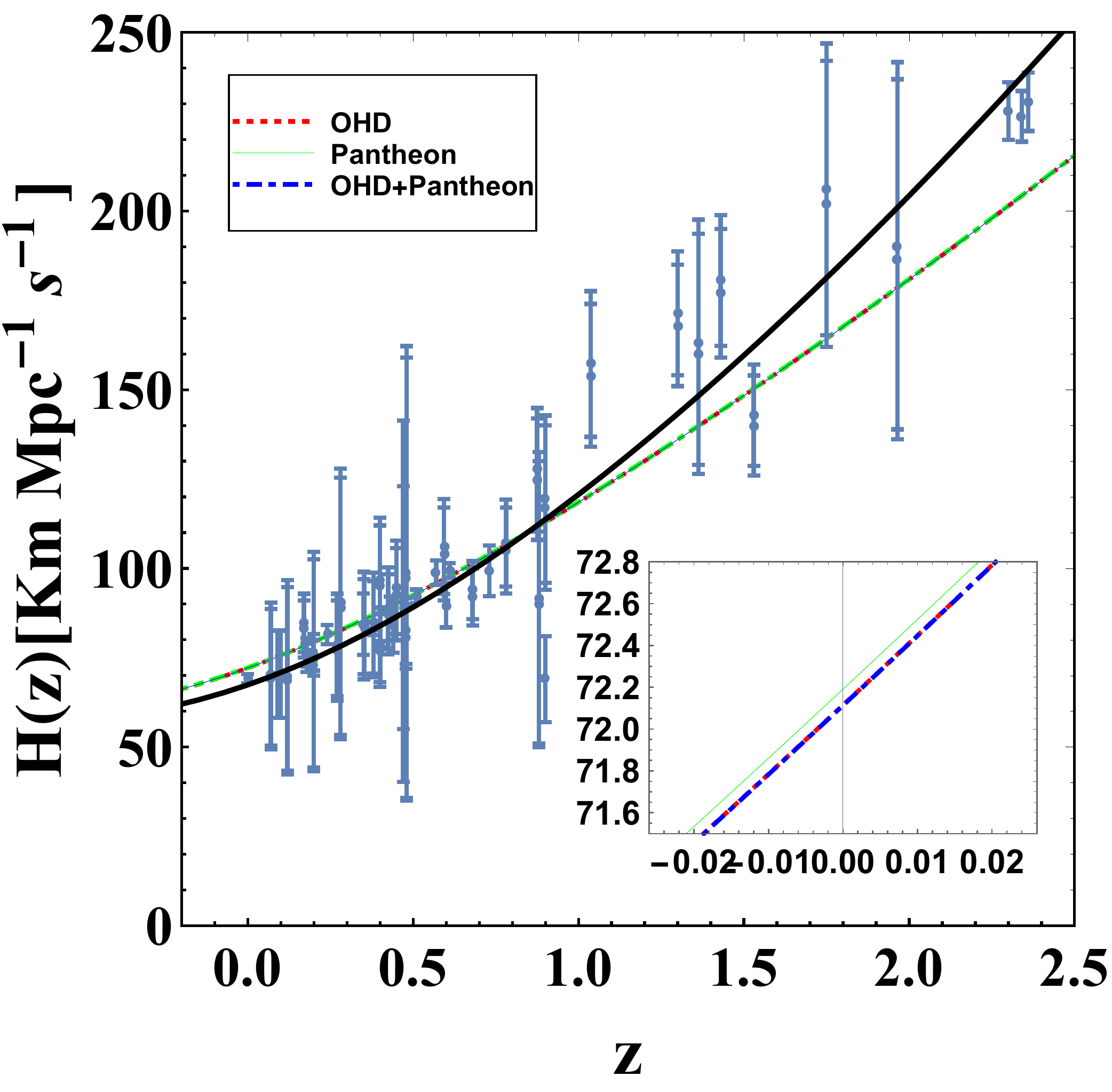}}\hfill
	\subfloat[]{\label{b}\includegraphics[scale=0.40]{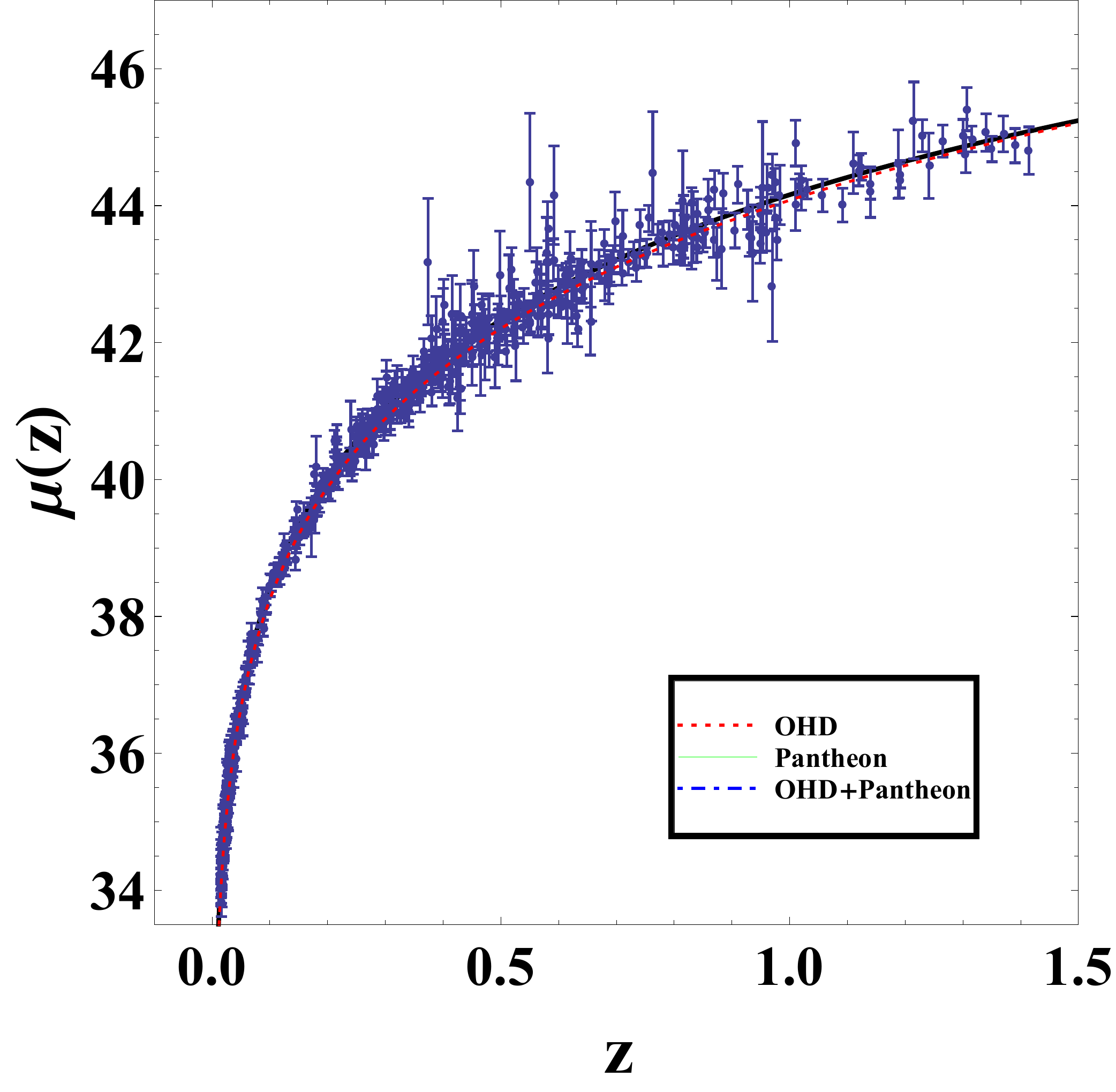}} 
	
\caption{\scriptsize The deviations of the model from $ \Lambda$CDM  are depicted by the graphs (a) and (b). The blue color error bars are used from the data $ OHD $ and $ SNeIa $. Various colored lines and black lines represent our obtained model and the $ \Lambda$CDM in the plots respectively.}
 \label{Fig:1}
\end{figure}
\begin{equation}\label{27}
H(z) = -\frac{1}{1+z} \frac{dz}{dt}.
\end{equation}

The constraints on $ n $ can be determined by $ \chi^2_{min} $. This is the same as performing a maximum likelihood analysis. The following is the definition of the probability function  $ \chi _{OHD}^{2}(n, H_0) $:
\begin{equation}\label{28}
\chi _{OHD}^{2}(n, H_0)=\sum\limits_{i=1}^{77}\left[ \frac{H_{th}(n,H_0,z_{i})-H_{obs}(z_{i})}{\sigma _{H(z_i)}}\right] ^2.
\end{equation}
Here, using $ z_{i} $, we calculate $ H(z_i) $. The observational Hubble dataset is represented by $ OHD $.  Here $ H_{th} $ represents the theoretical value and $ H_{obs}$ denotes the observed values of $ H $ in our model. The standard error is given by  $ \sigma_{H(z_{i})} $ in the observed value of $ H $.\\  

The model parameter $ n $ was restricted by the use of $ Pantheon $ bin data (48 points), which is the latest compilation dataset. These results are compared to the $ \Lambda$CDM values. The Chi-square function $ \chi_{OPN}^{2}(n, H_0) $ is given by   
\begin{equation}\label{29}
\chi _{OPN}^{2}(n, H_0)=\sum\limits_{i=1}^{48}\left[ \frac{\mu_{th}(n,H_0,z_{i})-\mu_{obs}(z_{i})}{\sigma _{\mu(z_{i})}}\right]^2.
\end{equation}

Here $ OPN $ belongs to the bin points dataset. The model's theoretical and observed distance modulus are $ \mu_{obs} $ and $ \mu_{th} $ respectively. $ \sigma_{\mu(z_{i})} $ represents the standard deviation of the observed value. In addition, $ \mu(z) $ represents the distance modulus.    
\begin{figure}
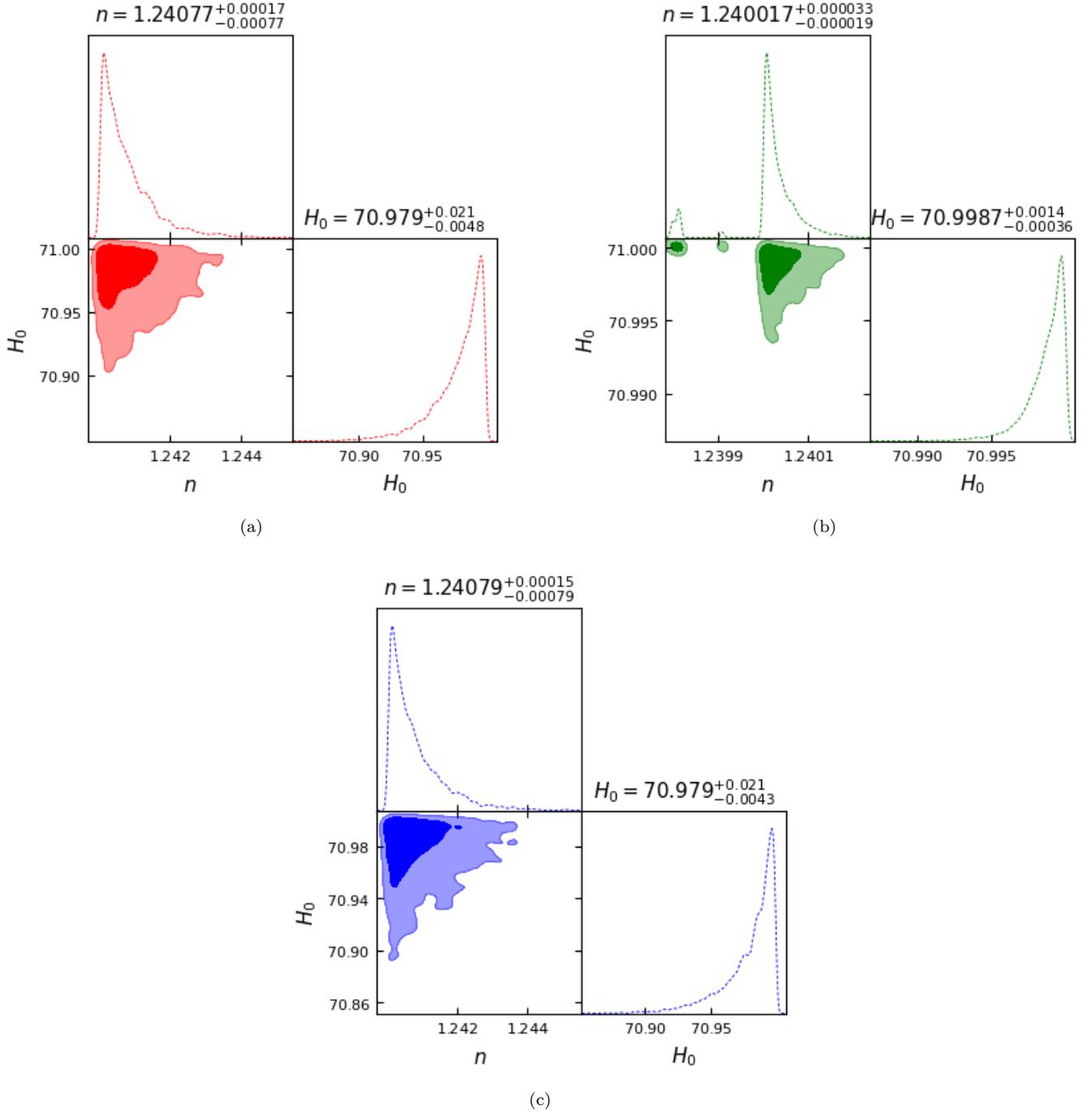
\centering
  \subfloat[]{\label{a}\includegraphics[scale=0.85]{HzC.pdf}}\hfill
  \subfloat[]{\label{b}\includegraphics[scale=0.85]{PnC.pdf}}\par
  \subfloat[]{\label{c}\includegraphics[scale=0.85]{HzPnC.pdf}} 
	
\caption{\scriptsize Figs. (a),(b) and (c) illustrate the $ n-H_0 $ likelihood contours for the data $ OHD $, $Pantheon$ bin data and their joint dataset respectively. The $1 \sigma $ and $ 2 \sigma $ errors are represented by the dark-shaded zone, the light-shaded region, and by extremely light-shaded region respectively.}
 \label{Fig:2}
\end{figure}
\begin{equation}\label{30}
\mu(z)= m-M = \mu_{0}+5Log D_l(z).
\end{equation}
A standard candle's apparent and absolute magnitudes are represented by $ m $ and $ M $, respectively. The parameters $ \mu_0 $ ( nuisance parameter) and $ D_l(z) $ (luminosity distance) for a flat universe are:
\begin{equation}\label{31}
D_l(z)=(1+z)H_0\int_0^z \frac{1}{H(z^*)}dz^*,
\end{equation}
and
\begin{equation}\label{32}
\mu_0= 5Log\Big(\frac{H_0^{-1}}{1Mpc}\Big)+25.
\end{equation}
We integrate $ OHD $ and Pantheon bin data to gain more compact restrictions on the model parameter $ n $ and to avoid dissipation in the observational data. As a result, we define
\begin{equation}\label{33}
\chi_{HP}^{2}=\chi _{OHD}^{2}+\chi _{OPN}^{2},
\end{equation}
\begin{table}[H]
\caption{ Summary of the numerical values.}
\begin{center}
\label{tab1}
\begin{tabular}{l c c c r} 
\hline\hline
\\ 
{Dataset} &     ~ Hubble parameter  & ~  Model parameter 
\\ 
 &      $ H_0 $ {\footnotesize(km/s/Mpc)}    &     $ n $  
\\
\\
\hline      
\\
{$ OHD $ }   &  ~ $ 70.979^{+0.021}_{-1.0048}  $   &  ~ $ 1.24077^{+0.00017}_{-0.00077} $  &   
\\
\\
{$ Pantheon $ }   &  ~ $ 70.9987^{+0.0014}_{-0.00036} $   &  ~ $ 1.240017^{+0.000033}_{-0.000019} $  
\\
\\
{$ OHD $ + $ Pantheon $ } &  ~$  70.979^{+0.021}_{-0.0043}  $  & ~ $ 1.24079^{+0.00015}_{-0.00079} $  
\\
\\ 
\hline\hline  
\end{tabular}    
\end{center}
\end{table}
The likelihood contours of $ n $ and $ H_0 $ in the plane $ n $-$ H_0 $ are presented in Fig.  \ref{Fig:2} with the errors $ 1 \sigma $ and $ 2 \sigma $. Such type of contours converge and may check using the Gelman-Rubin convergence test \cite{Gelman:1992zz, Singh:2022nfm}. The best-fit values with $ 1 \sigma $ error of $ n $ and $ H_0 $ are constrained and are placed in Table \ref{tab1}. 

\begin{figure}\centering
	\subfloat[]{\label{a}\includegraphics[scale=0.45]{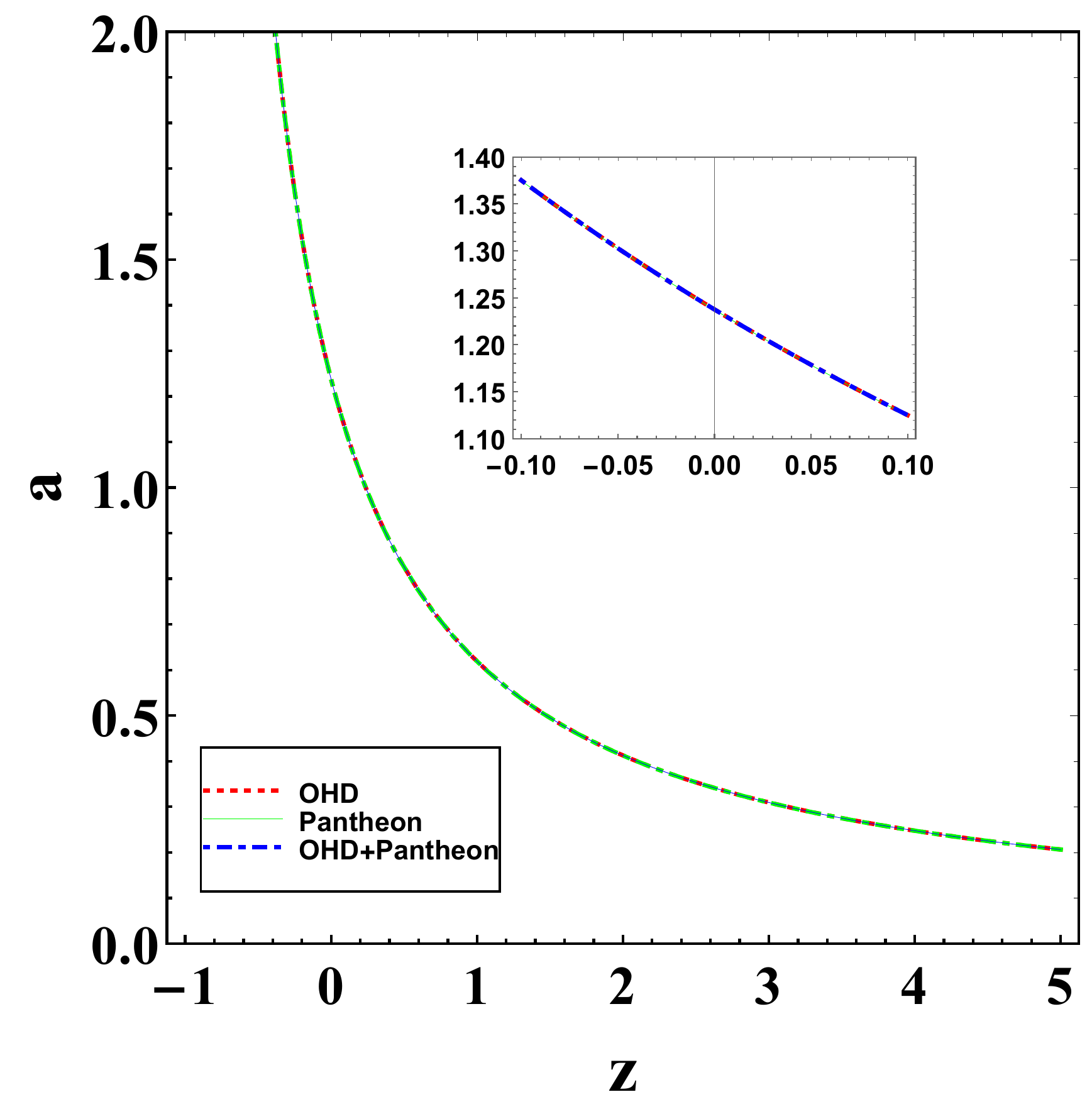}}\hfill
	\subfloat[]{\label{b}\includegraphics[scale=0.45]{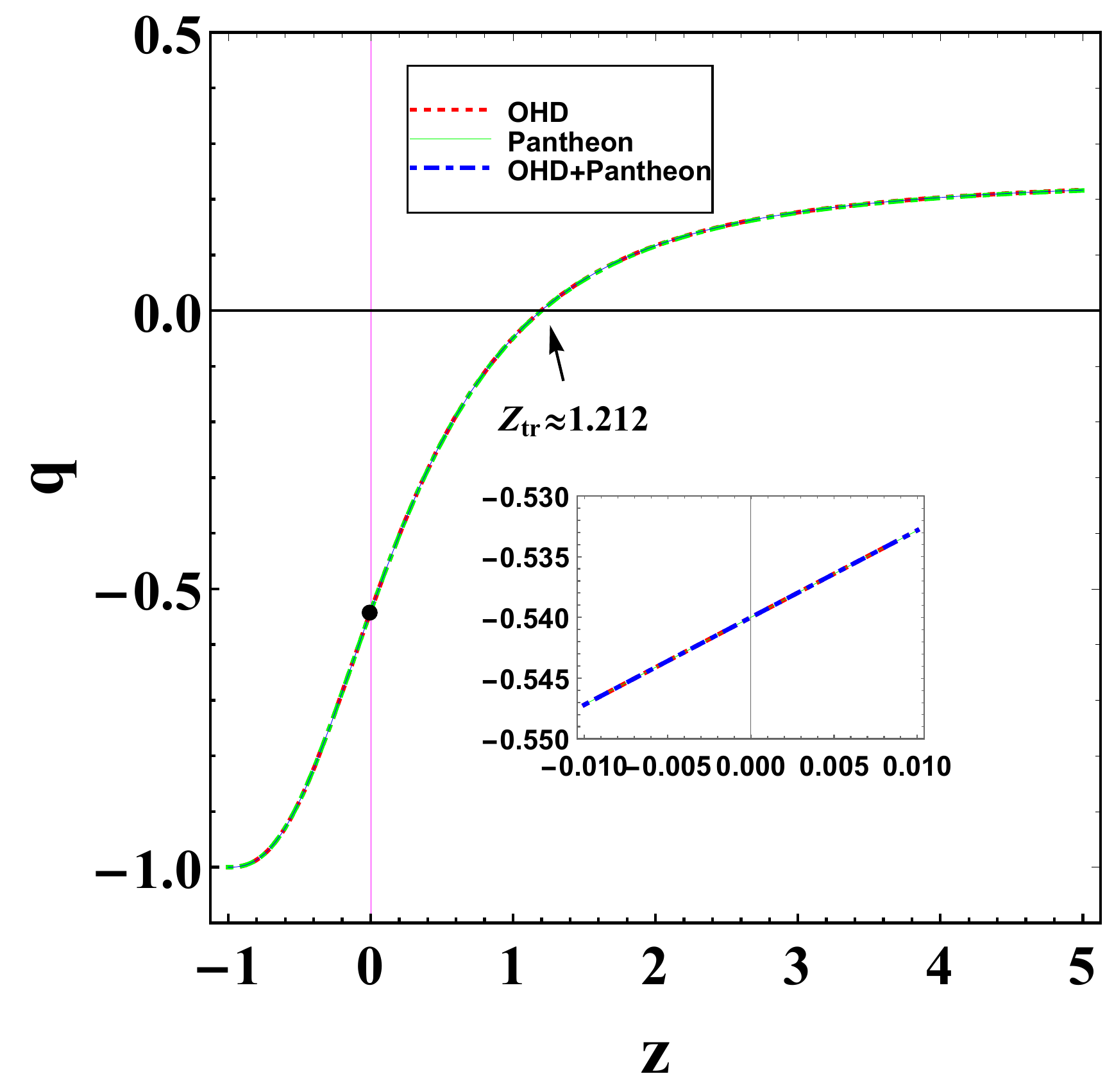}}\par
	\subfloat[]{\label{c}\includegraphics[scale=0.45]{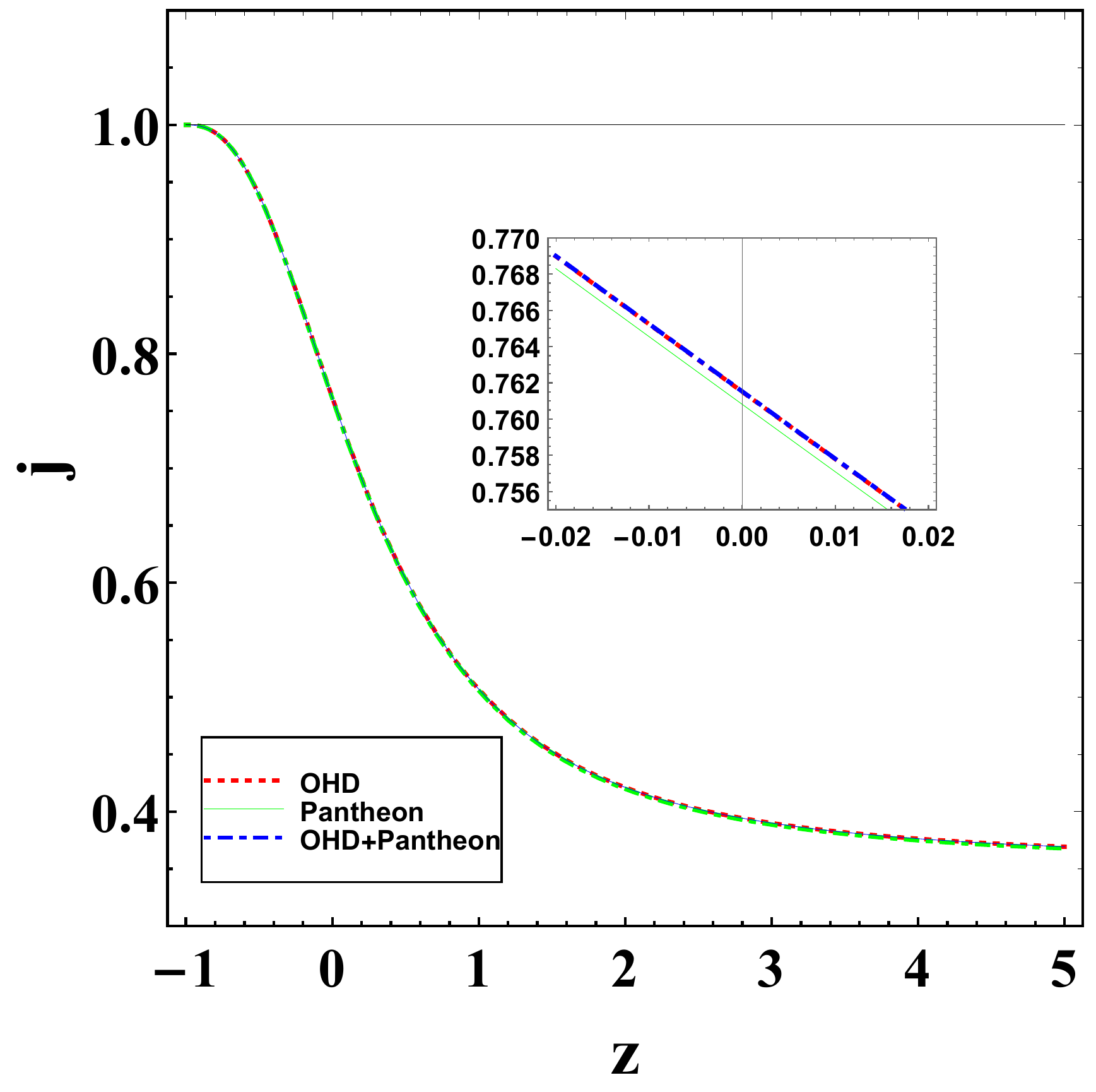}}
\caption{\scriptsize The variations of scale factor, deceleration parameter, and jerk parameter \textit{w.r.t.} $ z $.}
 \label{Fig:3}
\end{figure}

\section{Physical Interpretation of the model}
 
\qquad Without loss of any generality we can assume $ \beta=H_0 $. The most recent value of the Hubble parameter and DP are $ 67.4 Km/s/Mpc $ and $ q_0=-0.54 $ respectively. The values of $ H_0 $ and $ q_0 $ are taken by most recent Plank results \cite{Planck:2018vyg}. The scale covariant parameter $\alpha $ is fixed as $ 1.7 $, $\phi_0$ is fixed as $1$ and the model parameters $ \beta $ and $ n $ have been constrained using $ OHD $ data of 77 data points, $ Pantheon $ data of 48 bin points and joint $ OHD + Pantheon $ data. With these constraint values, we plot the curves for $ H $ and $ q $ versus redshift $ z $ by using Eqs. (\ref{20}) and (\ref{22}) (see Fig. \ref{Fig:3}).\\
\begin{figure}\centering
	\subfloat[]{\label{a}\includegraphics[scale=0.50]{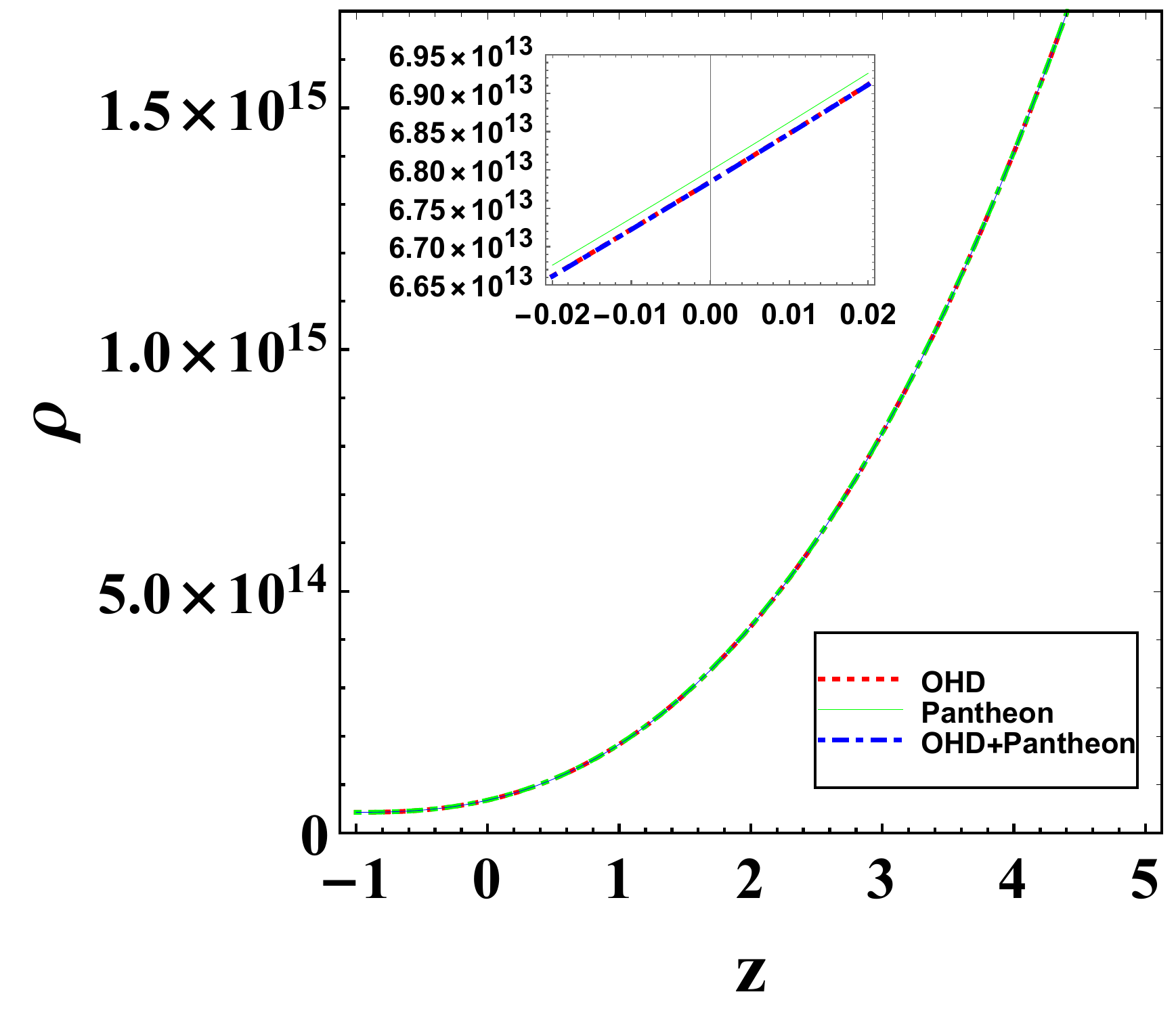}}\hfill
	\subfloat[]{\label{b}\includegraphics[scale=0.50]{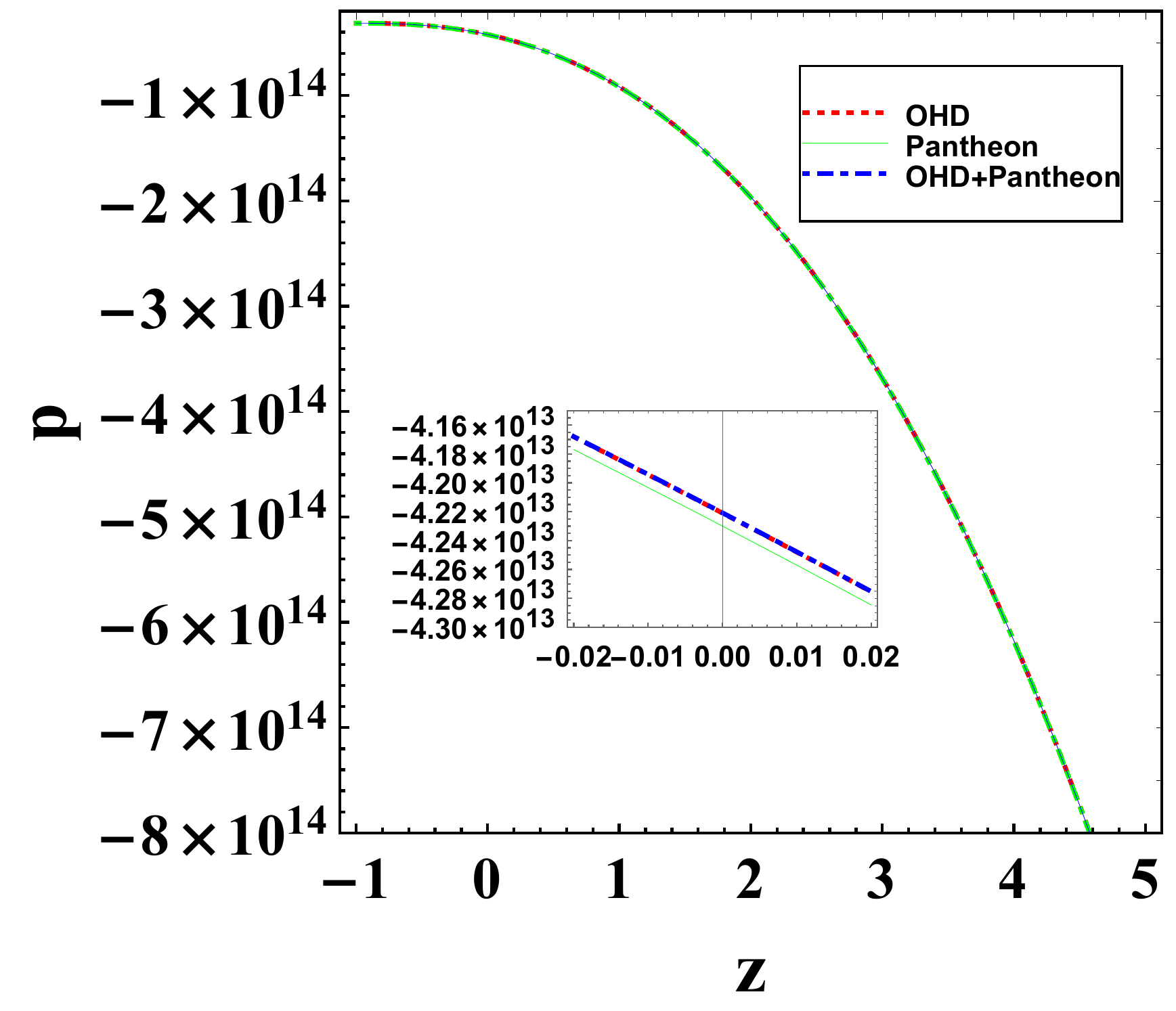}}\par
	\subfloat[]{\label{c}\includegraphics[scale=0.50]{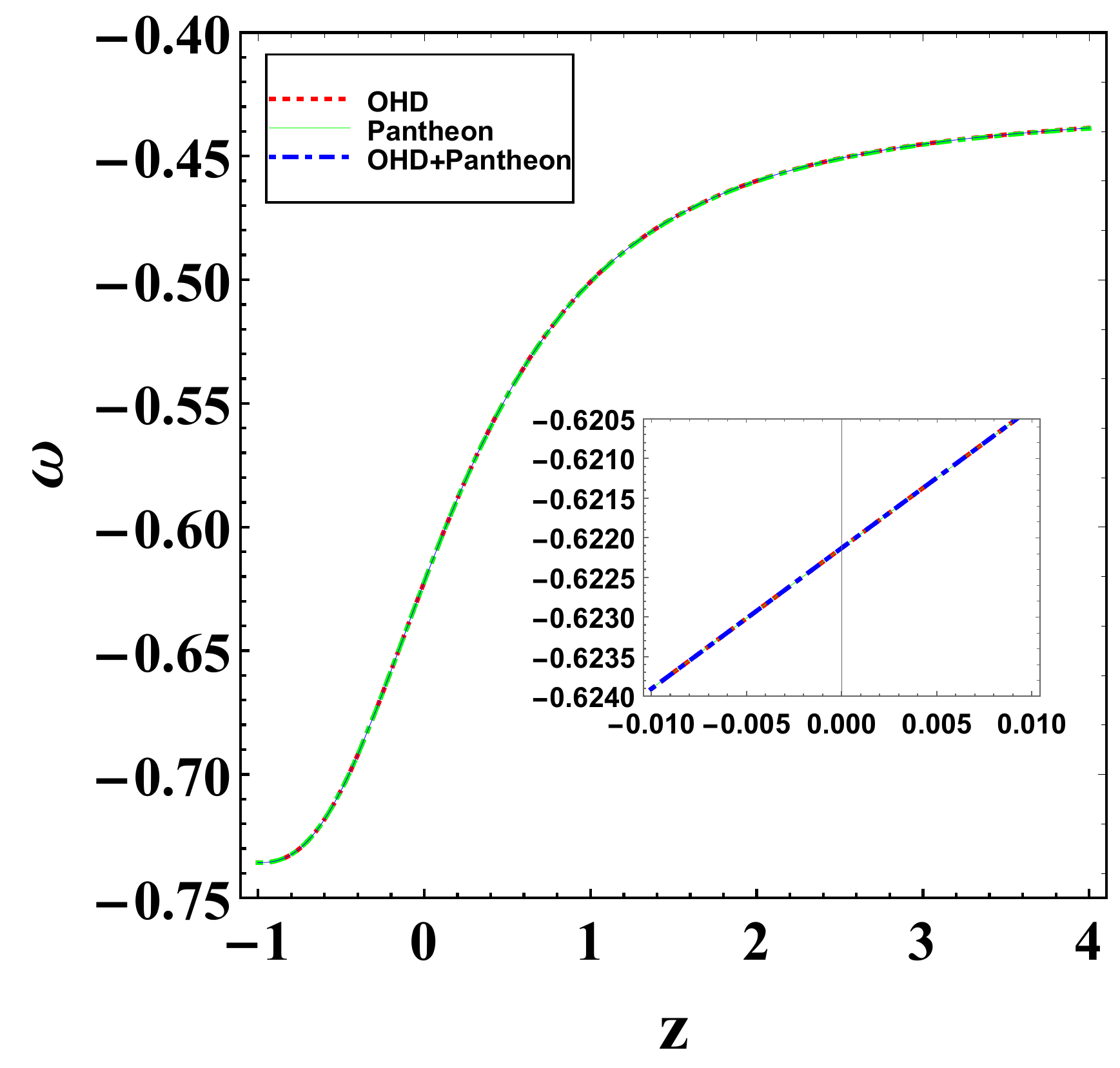}}
\caption{\scriptsize The variations of density parameter, pressure and EoS-parameter \textit{w.r.t.} $ z $.}
 \label{Fig:4}
\end{figure}
The scale factor is increasing from early to the late universe and the present value of $ a $ in this model is 1.237 approximately (see Fig. \ref{Fig:3}a). Also in Fig. \ref{Fig:3}b, we see deceleration at the beginning of the Universe's evolution and acceleration at the present age ($ z=0 $) as well as at later times in our model. In this model, the present value of DP is approx $ -0.54 $ for all the above observational data. The transition points from deceleration to acceleration are given by $ z_{tr}\simeq 1.212 $ for all datasets. Thus, the Universe is accelerating and the expansion rate is dropping very slowly. 

To examine the geometrical nature of the Universe we also study the parameters which contain the higher derivatives of the scale factor. These parameters are known as cosmographic parameters and are defined as:
\begin{equation} \label{37}
~~j=\frac{1}{aH^{3}} \frac{d^3 a}{dt^3}%
,~~s=\frac{1}{aH^{4}} \frac{d^4 a}{dt^4},
~~l=\frac{1}{aH^{5}} \frac{d^5 a}{dt^5},
~~m=\frac{1}{aH^{6}} \frac{d^6 a}{dt^6}.
\end{equation}
These parameters are separately known as jerk, snap, lerk, and m parameters. Now to understand the behavior of the cosmographic parameters, we draw the trajectories for the various observational datasets. In Fig. 3(c), we observe that the jerk parameter $ j $ converges to $ 1 $ in late times, which reveals that our model behaves like $ \Lambda $CDM in late times for all observations. 

\begin{figure}\centering
	\subfloat[]{\label{a}\includegraphics[scale=0.43]{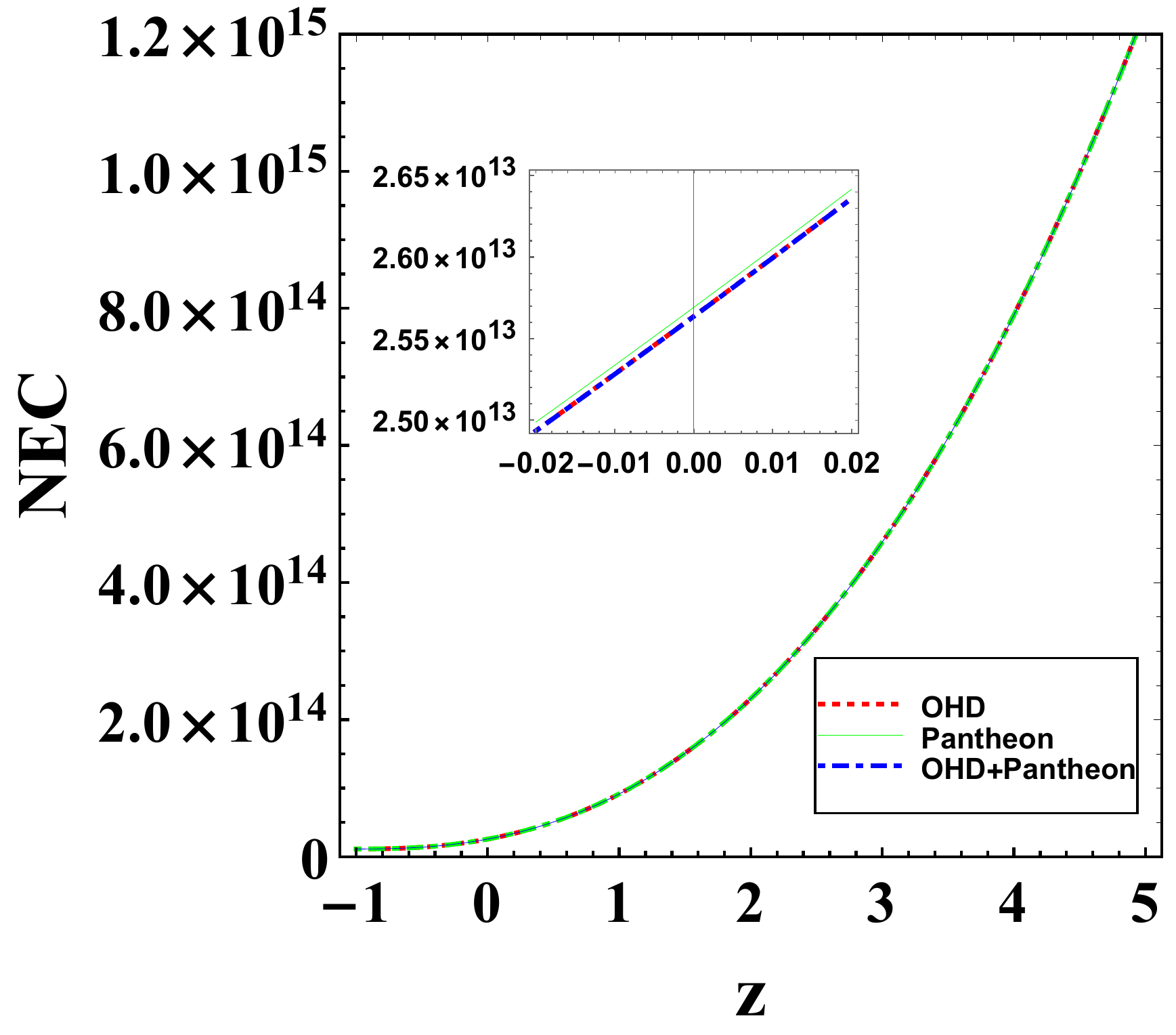}}\hfill
	\subfloat[]{\label{b}\includegraphics[scale=0.43]{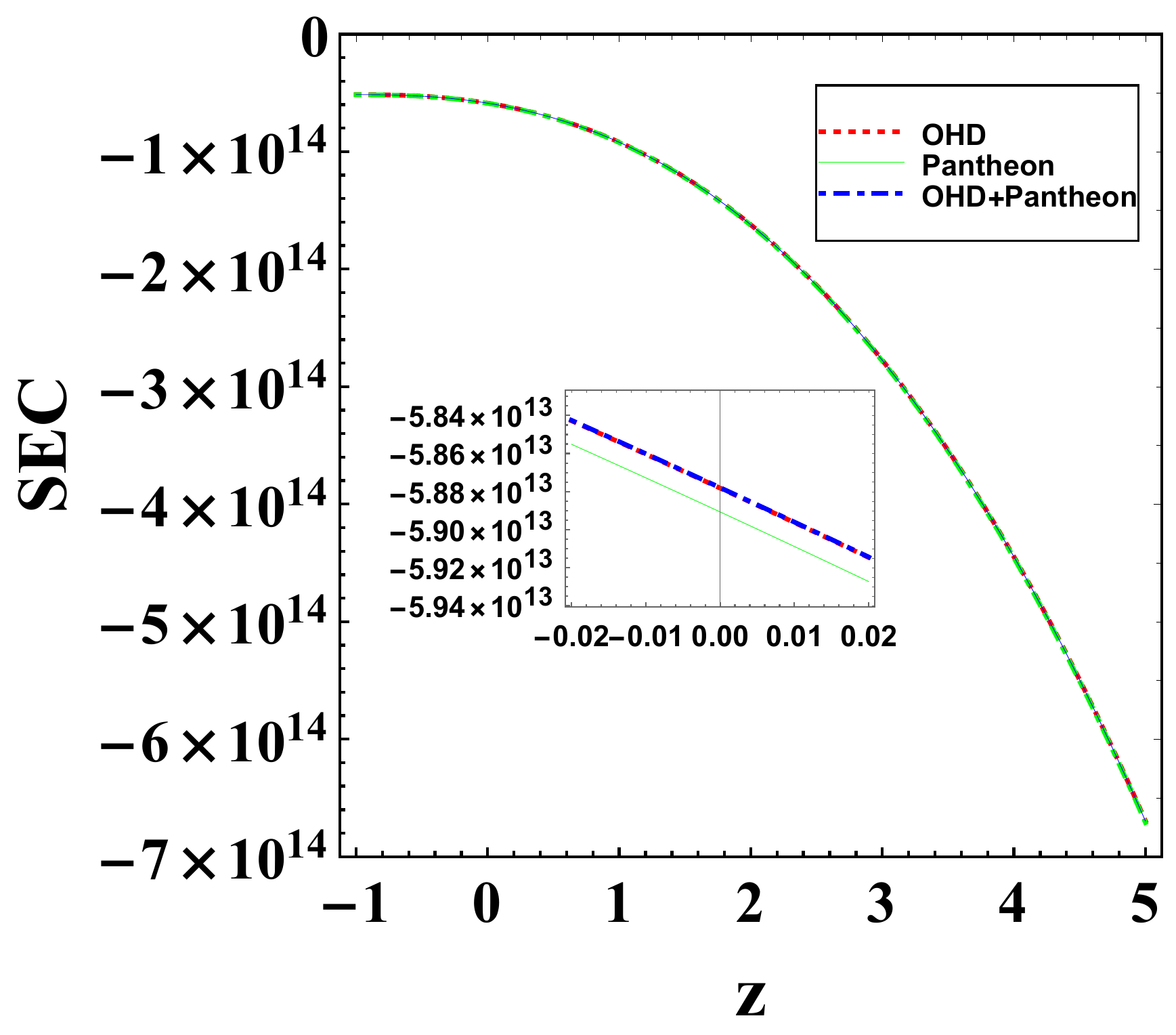}}\par 
	\subfloat[]{\label{c}\includegraphics[scale=0.43]{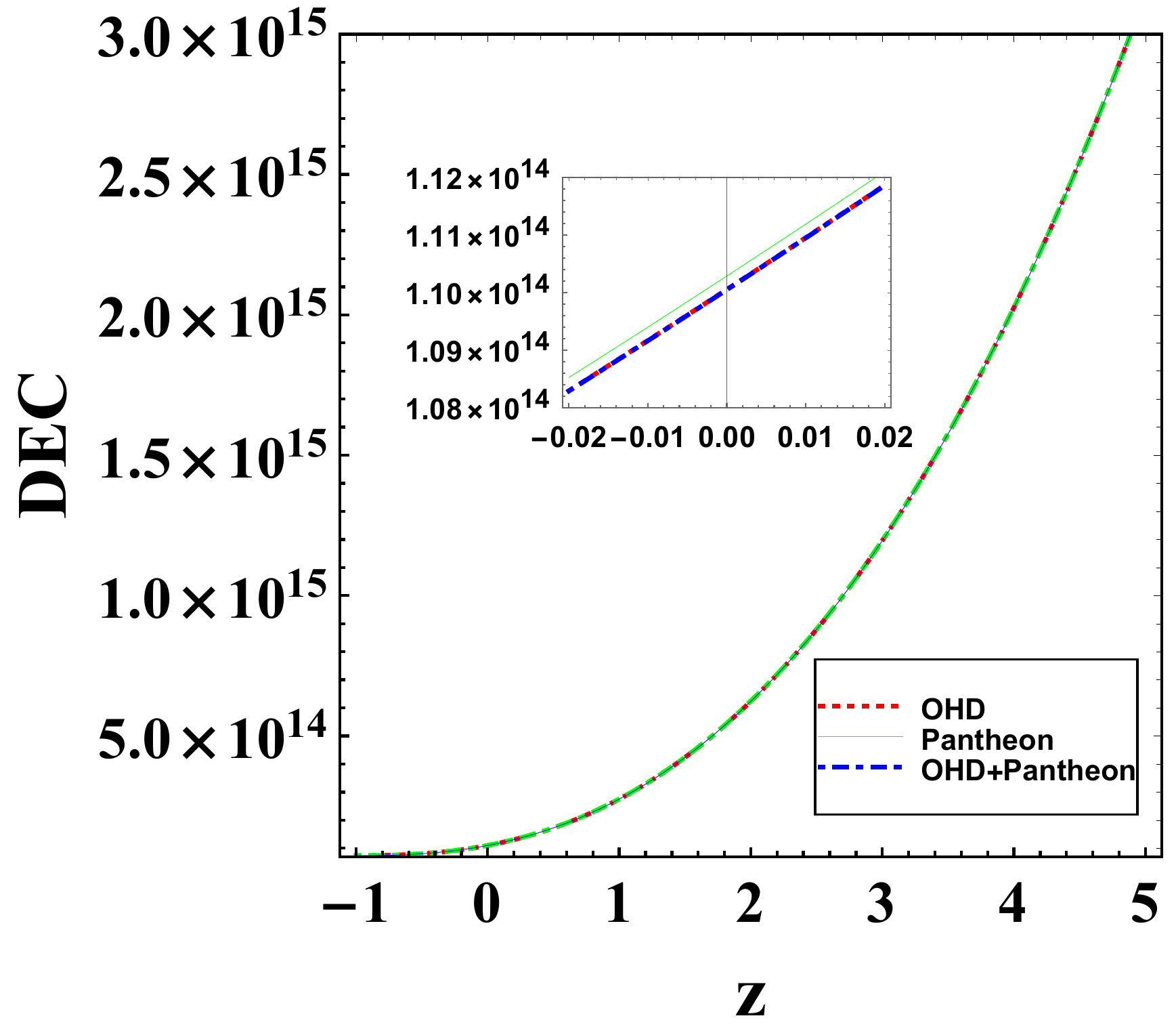}}
	\caption{\scriptsize The plots for the various energy conditions.}
	\label{Fig:5}
\end{figure}

According to the data $ OHD $, $ Pantheon $  and $ H(z)+ Pantheon $, the slope of $ \rho $ $ - $ $ z $ plot reduces and as a result, we find that the energy density of the model $ \rho $ is quite large in the beginning during the Universe's evolution and then gradually decreases, indicating that the amount of energy density is monotonically decreasing in late periods. The behavior of the curve of energy density can be viewed in Fig. \ref{Fig:4}a. In Fig. \ref{Fig:4}b it is seen that isotropic pressure $ p $ is highly negative initially and afterward, it increases but remains negative for a whole range of z. At present ($ z=0 $) pressure is negative which shows the existence of dark energy in this model. The conduct of the EoS parameter $ \omega $ explores the many phases of the cosmic history of the Universe, as shown in Fig. \ref{Fig:4}c. This model shows the quintessence model in early times, at present as well as in late times. Thus, we conclude that this model shows a quintessence dark energy model at the present age.

In General Relativity (GR), various energy conditions (ECs) are discussed in GR to describe the feasible concept of the fact $\rho$ can never be negative in the Universe to the entire EMT \cite{Santos:2005pe, Santos:2007zza, Sen:2007ep, Singh:2010zze, Singh:2010zzo, Mandal:2022ata}. Many ECs are routinely used in GR, and their viability may be determined using the well-known Raychaudhuri equation \cite{Padmanabhan:2002vv, carr}. The energy conditions are commonly presented in one of two ways either geometrically, where the energy conditions are given because of the Ricci or Weyl tensors, or physically, where ECs are stated either in terms of $T_{ij}$ or in $ \rho $. The null energy condition (NEC), weak energy condition (WEC), strong energy condition (SEC), and dominant energy condition (DEC) are given by $ \rho+p \geq 0 $; $ \rho \geq 0 $, $ \rho+p \geq 0 $; $ \rho+ p \geq 0 $, $ \rho+3p \geq 0 $; and $ \rho \geq |p| $ respectively. 

All the energy conditions depicted in Fig. \ref{Fig:5}, where we observe that NEC and DEC are satisfied but SEC is violated in our model\cite{Bolotin:2015dja, Visser:1997qk, Visser:1997tq, Singh:2022eun, Singh:2022wwa}.

\begin{figure}\centering
	\includegraphics[scale=0.40]{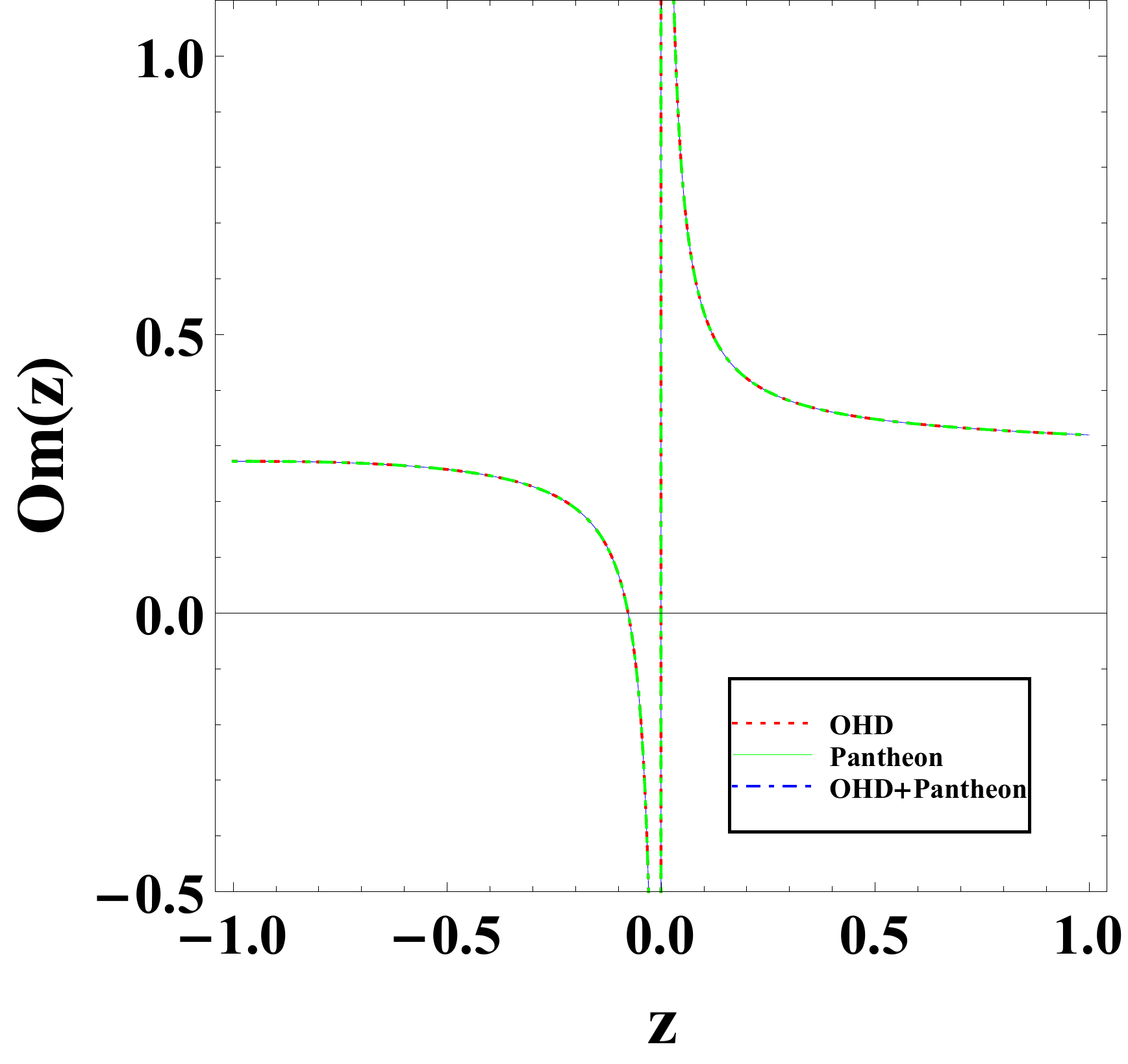}
\caption{\scriptsize The variation of $ Om(z) $ \textit{w.r.t.} $ z $.}
\label{Fig:6}
\end{figure}

\section{Cosmographic Analysis}

In this section, various cosmographic parameters are studied, which are useful to discriminate between various dark energy models. \textbf{Sahni et al.} \cite{Sahni:2008xx} discussed a model-independent diagnostic parameter, known as $ Om(z) $. This helps us to differentiate various DE models from $ \Lambda $CDM and depends only on $ H $. It is formulated as
\begin{equation}\label{34}
Om(z)=\frac{(\frac{H}{H_0})^2 -1}{(1+z)^3 -1},
\end{equation}
which is calculated as 
\begin{equation}\label{35}
Om(z)=\frac{\frac{0.00022013 \beta ^2 (\text{q0}+1) (z+1)^{2 n} (\frac{(n-\text{q0}-1) (z+1)^{-2 n}}{\text{q0}+1}+1)}{n^2 (n-\text{q0}-1)}-1}{z^3+3 z^2+3 z}.
\end{equation}

With the help of $ Om(z)~vs. ~z $ plot we analyze the behavior of the dark energy model. The trajectory of $ Om(z) $ with a negative curvature represents the quintessence model and with a positive curvature shows the phantom model. Also, if the curvature of $ Om(z) $ is zero then the model is similar to the $ \Lambda $CDM model. Here, in Fig. \ref{Fig:6}, the trajectories of $ Om(z) $ show the negative slope in late times. Hence, our model exhibits quintessence behavior for all the observational data in late times. 

Next, geometrical analysis of our models is done along different approaches. Sahni et al. \cite{Sahni:2002fz}
and Alam et al. \cite{Alam:2003sc} introduce two geometrical diagnostic parameters $ (r,s^*) $, these are known as statefinder parameters and defined as
\begin{equation}\label{36}
r=\frac{\dddot{a}}{aH^3} ~~~~~~ and ~~~~~~ s^*=\frac{r-1}{3(q-\frac{1}{2})}.
\end{equation}
\begin{figure}\centering
	\subfloat[]{\label{a}\includegraphics[scale=0.28]{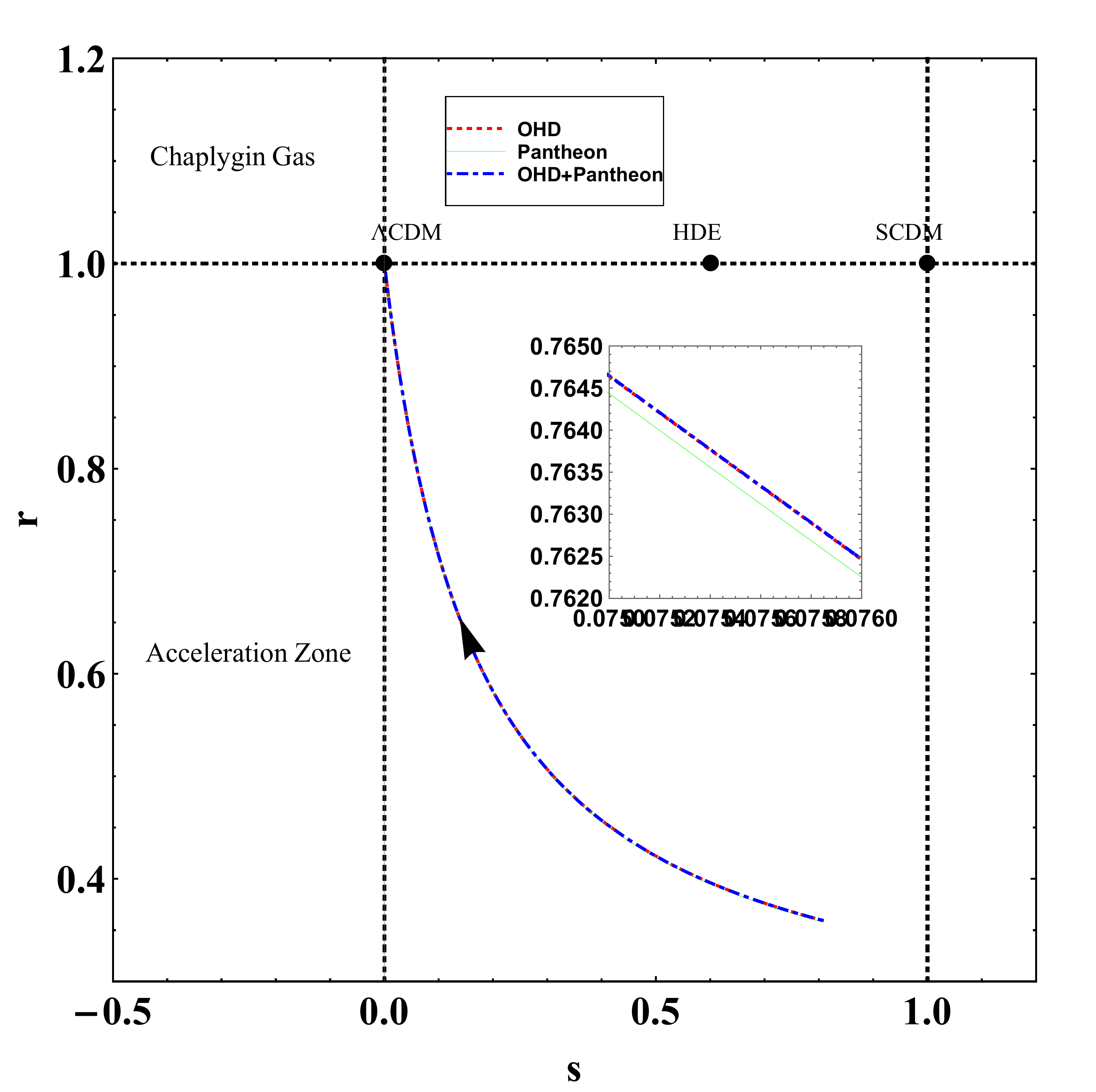}}\hfill
	\subfloat[]{\label{b}\includegraphics[scale=0.30]{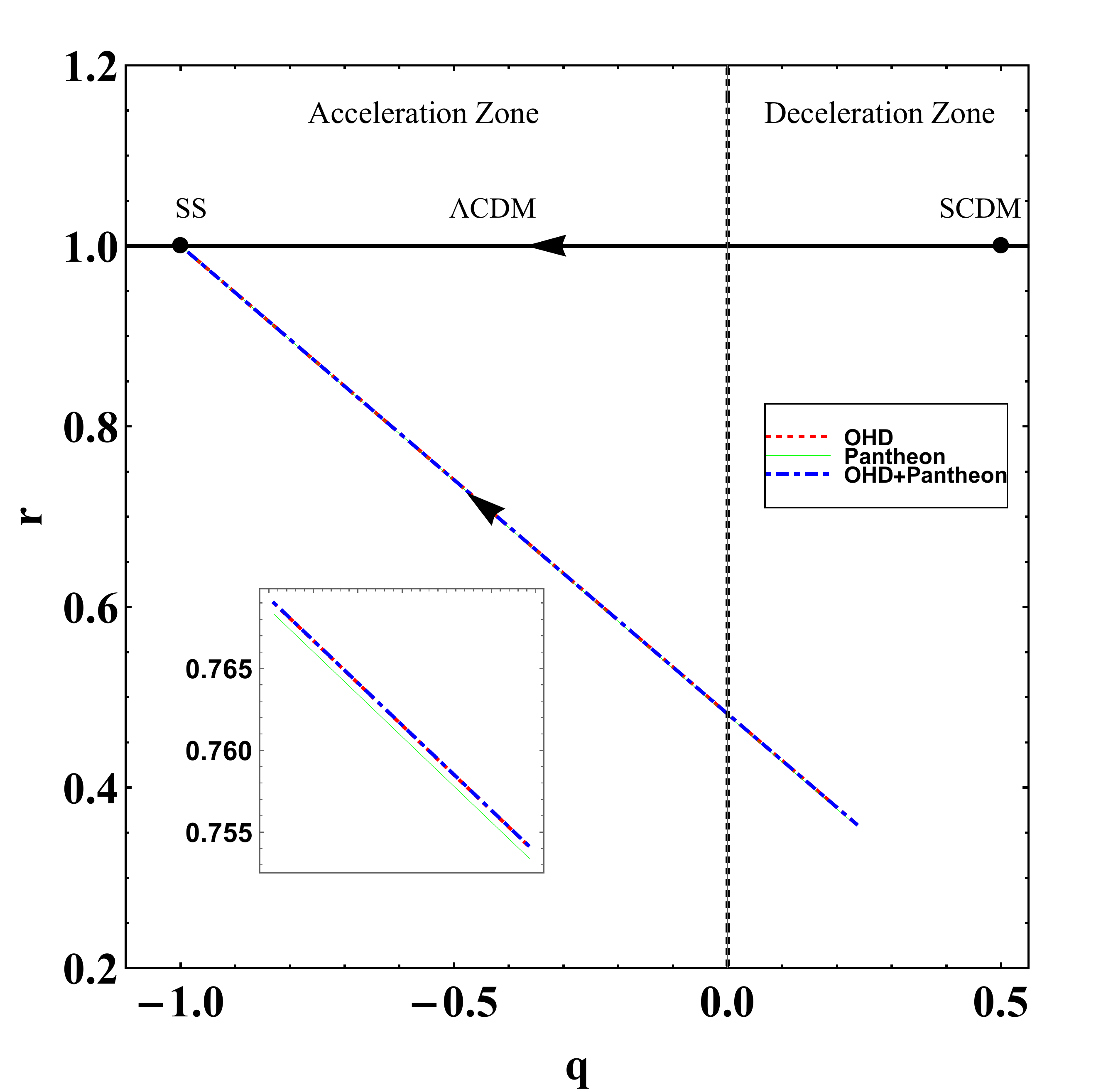}} 
\caption{\scriptsize The plots depict $ s^*- r $ and $ q-r $,}
\label{Fig:7}
\end{figure}
where $ q\neq \frac{1}{2} $. Statefinder diagnostic parameters are used to compare the goodness of various dark energy models with $ \Lambda $CDM. The points $ (r,s^*)=(1,0) $ and $ (r,s^*)=(1,1) $ represent $ \Lambda CDM $ model and $ SCDM $ (Standard cold dark matter) model respectively. The value of $ r $ and $ s^* $ is obtained from Eqs. (\ref{15}), (\ref{20}), (\ref{22}) and (\ref{36}). Here by using the best-fit values of parameters, we find these trajectories (see Fig. 7(a). And the arrows represent the direction of the evolution of the trajectories. In our model $ r<1 $ and $ s^*>0 $ at early time, which corresponds to the quintessence DE model. The present value of $ (r_0,s^*_0)=(0.762,0.076) $, $(r_0,s^*_0)=(0.761,0.077) $, $(r_0,s^*_0)=(0.762,0.076) $ for data $ OHD $, $ Pantheon $ and their joint respectively, which is near to $ (1,0) $. And in late time, our model collapsed to $ \Lambda $CDM model.

In the $ q-r $ plane, point $ (r,q)=(1,\frac{1}{2})$ represents the SCDM model whereas $ (1,-1) $ represents SS (steady state) model and the dotted horizontal line $ r=1 $ is for $ \Lambda $CDM model. In Fig. 7(b) trajectories are switching their value from positive to negative, which shows the phase transition from deceleration to acceleration. And all the trajectories approach $SS$ in the future.

\section{Thermodynamic Analysis}
\qquad According to the second law of thermodynamics, the total entropy of the universe increases with respect to time. Here, in this section, we study the thermodynamic part and for that total entropy of the universe is calculated. Let $ S $ be the total entropy which consists of both, the entropy of matter inside $ S_i $ and entropy on the boundary $ S_o $. Now let us assume that the boundary of the universe is bounded by the radius of the apparent horizon ($ r_h $), which is calculated by using the scale factor for the flat FLRW metric as \cite{Hawking:1975vcx, Singh:2019uwv, Brevik:2020psn}:
\begin{equation}\label{38}
r_h = \frac{a}{\dot{a}} = \frac{n}{\beta\,\coth\left(\sinh^{-1}\sqrt{\frac{n-(q_0+1)}{(z+1)^{2n}(q_0+1)}}\right)},
\end{equation}
\begin{figure}\centering
	\subfloat[]{\label{a}\includegraphics[scale=0.52]{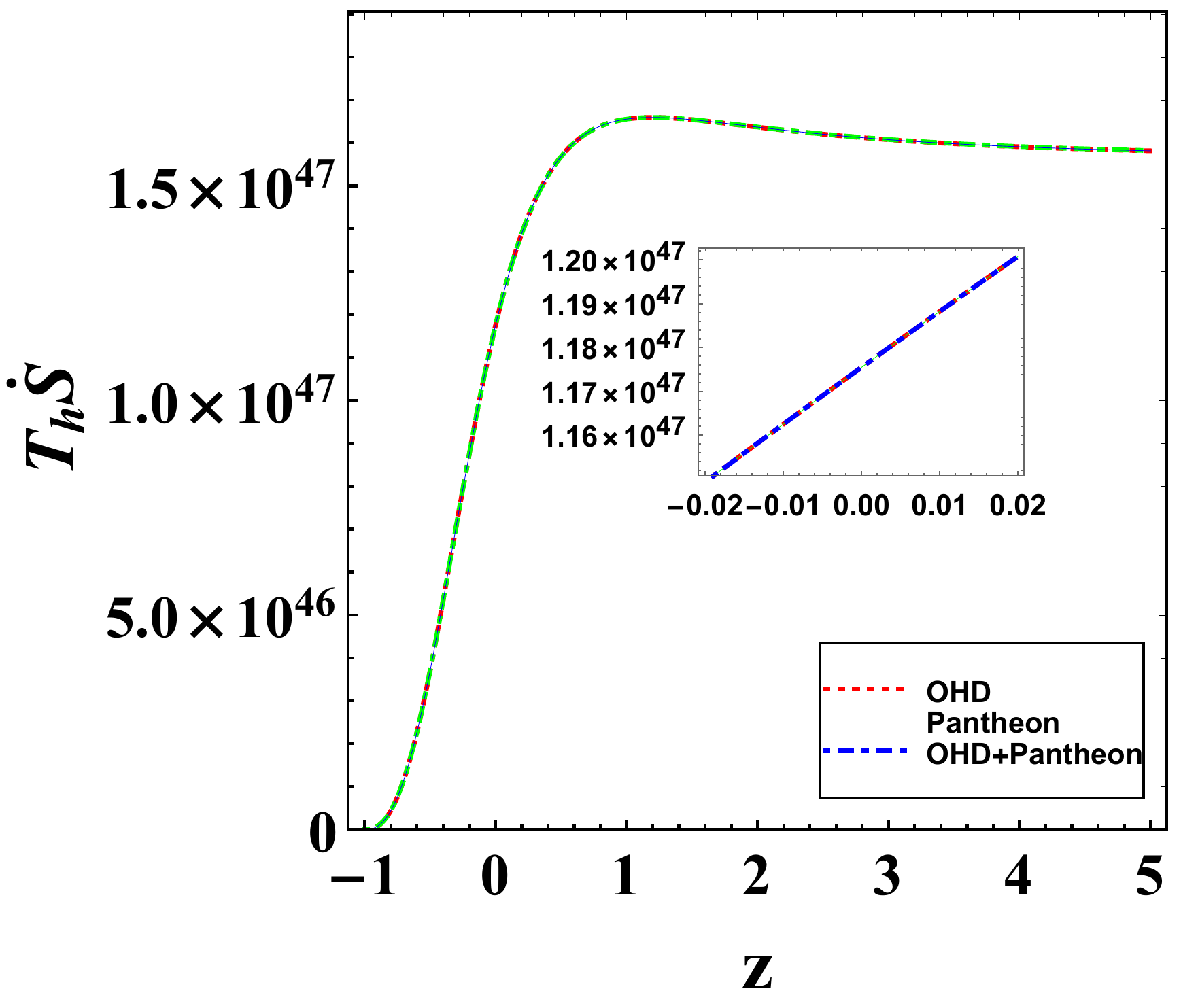}}\hfill
    \subfloat[]{\label{b}\includegraphics[scale=0.44]{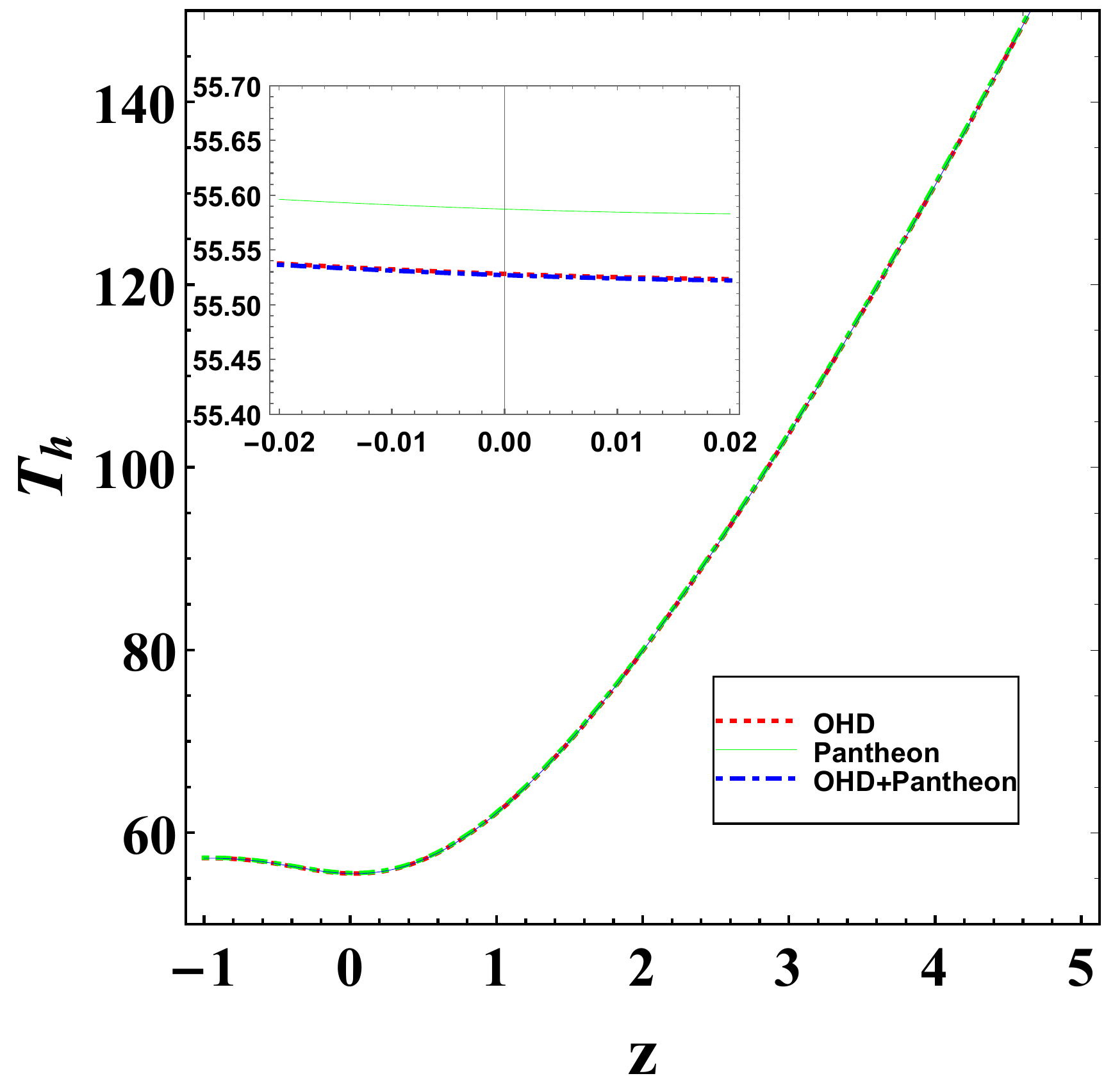}}
\caption{\scriptsize The variations of total entropy $ T_h \dot{S} $ and temperature $ T_h $ \textit{w.r.t.} $ z $.}
\label{Fig:8}
\end{figure}
and the entropy on the boundary of the horizon is
\begin{equation}\label{39}
S_o=\frac{\pi \kappa_B r_h^2}{l_{Pl}^2},
\end{equation}
where $ \kappa_B $ and $ l_{Pl} $ are  Boltzmann constant and Planck's length respectively. Also using (\ref{39}), derivative of $ S_o $ is obtained as
\begin{equation}\label{40}
\dot{S}_o=\frac{2 \pi \kappa_B n^2 \sqrt{\frac{(n-\text{q0}-1) (z+1)^{-2 n}}{\text{q0}+1}}}{l_{Pl}^2\beta  \left(\frac{(n-\text{q0}-1) (z+1)^{-2 n}}{\text{q0}+1}+1\right)^{3/2}}.
\end{equation} 
Here, the value of $ \dot{S}_o $ is positive for all observations. Now, to investigate the matter entropy inside the horizon we use the Gibbs relation 
\begin{equation}\label{41}
T_h dS_i = d(\rho V) + p dV = V d(\rho) + (\rho + p) dV,
\end{equation}
where $ V=\frac{4\pi}{3} r_h^3 $ is the volume enclosed by the horizon and $ T_h $ is the Hawking temperature on the boundary of horizon.
\begin{equation}\label{42}
T_h=\frac{\beta  \sqrt{\frac{(n-\text{q0}-1) (z+1)^{-2 n}}{\text{q0}+1}+1} \left(1-\frac{n}{2 \left(\frac{(n-\text{q0}-1) (z+1)^{-2 n}}{\text{q0}+1}+1\right)}\right)}{n \sqrt{\frac{(n-\text{q0}-1) (z+1)^{-2 n}}{\text{q0}+1}}}.
\end{equation}
Now,  differentiating Eq. (\ref{41}) \textit{w.r.t.} time, we obtain that $ \dot{S_i}\geq 0 $. It is well known that the total entropy should not decrease concerning the time evolution \textit{i.e.}
\begin{equation}\label{50}
\dot{S}= \dot{S}_i + \dot{S}_o \geq 0.
\end{equation}
Fig. \ref{Fig:8}a exhibits that the total entropy $ T_h \dot{S} $ is positive in the entire redshift range of $ z $. It increases exponentially in the higher redshift range of $ z $ and decreases abruptly in late times. In Fig. \ref{Fig:8}b, it is visible that the Hawking temperature $ T_h $ is positive in the entire redshift range of $ z $ and decreases in late times. Thus, we see that the second law of thermodynamics corroborates our model for all observations.

\section{ Conclusion}

\qquad We investigate the functioning of the model in the Scale-covariant theory of gravity \cite{Canuto:1977zz} based on a flat FLRW metric. The presence of the scalar field affects the equation of continuity of the matter field. We start our solution by taking the deceleration parameter as a function of $ t $ and we get a hyperbolic solution of scale factor $ a(t)=\sinh^{1/n}(\beta t) $, where $ n $ and $ \beta $ are both positive constants. The error bar plots in Fig.\ref{Fig:1} show the deviation of our model with $ \Lambda $CDM. The constrained values of model parameters $ n $ and $ \beta=H_0 $ are given in Table \ref{tab1} and the likelihood contours are represented in Fig.\ref{Fig:2} for all observations. We evaluate the constrained values of Hubble parameter $ H_0=70.979^{+0.021}_{-0.0043} $ and the model parameter $ n=1.24079^{+0.00015}_{-0.00079} $ using joint analysis of the $ OHD $ data of 77-points and Pantheon bin data of 48 points. Performing the grid search in the whole parametric space ($  H_0>65 $, $ n>1  $), we find that the constrained values of $ H_0 $ which is closer to the recent Planck's data. Also, the values of deceleration parameter $ (q_0) $ are consistent and the other physical parameters are also given more compatible results which motivate us to work in the scale covariant theory.

This model starts with point-type singularity because it begins with a point of zero volume, infinite energy density, and temperature. The model has the behaviour of the late-time universe, which is the ever-accelerated expansion and faces a Big Freeze at the end. The model shows the quintessence dark energy era from high redshift to low redshift. The present value of the deceleration parameter is computed $ (q_0) \approx {-0.54} $. The transition point from deceleration to acceleration is given by $ z_{trj}\approx1.212 $ by using joint data analysis (see Fig. \ref{Fig:3}). The jerk parameter is consistent with $ \Lambda $CDM in late times $ \forall $ observations. The energy density $ (\rho) $, isotropic pressure $ (p) $ and the EoS parameter $ (\omega) $ are depicted. As a result, we observe that during the evolution of the universe, the energy density of the model $ \rho $ is initially relatively high and then gradually drops, showing that the amount of energy density is monotonically decreasing in late times. Isotropic pressure is negative for a whole range of z and shows dark energy at present as well as in the future. The behaviour of the EoS parameter $ \omega $ has been examined at several stages of cosmic evolution (see Fig. \ref{Fig:4}).

Several energy conditions (ECs) of the model are discussed to describe the feasible concept of the fact that the energy density can never be negative in the universe to the entire EMT. In Fig. \ref{Fig:5}, the energy conditions are in favour of quintessence dark energy. The energy conditions NEC and DEC are satisfied whereas SEC is violated. In Fig. \ref{Fig:6}, the trajectories of $ Om(z) $ exhibit the quintessence behaviour of the model for all the observational data in late times. 

The $ s^*-r $ trajectories deviate from $ SCDM $ traverse from the quintessence region and converge to $ \Lambda $CDM in late times. The $ q-r $ trajectories $ \forall $ observations evolve from the decelerating zone in early times to the accelerating zone and converge to $ SS $, the Steady State in late times (see Fig. \ref{Fig:7}). In addition, we obtain the total entropy is positive from early to late times which shows the validation of the second law of thermodynamics. The Hawking temperature $ T_h $ is positive in the entire redshift range of $ z $ and decreases as we move from the early Universe to the late Universe (see Fig. \ref{Fig:8}). Finally, we conclude that our model is an ever-expanding accelerating model and intended to Big Freeze probably at the end and shows a quintessence model in late times. 
\vskip0.2in 
\textbf{\noindent Acknowledgements} 
Shaily gratefully acknowledges Prof. J. P. Saini, Hon’ble Vice Chancellor, NSUT, New Delhi for the fellowship under the TRF scheme. JRLS would like to thank CNPq (Grant no. 309494/2021-4), and PRONEX/CNPq/FAPESQ-PB (Grant nos. 165/2018, and 0015/2019) for financial support. The authors express their sincere thanks to Prof. H. Parthasarathy, ECE, NSUT, New Delhi, India for fruitful discussions. The authors also express their thanks to the referee for his valuable comments and suggestions.
\vskip0.2in
\textbf{Appendix:} \\
\textbf{Remark for action principle:}\\

Even if the matter field is not coupled to the external scalar field, the equation of continuity for matter current will contain terms involving the external scalar field indeed the total action for the gravitational, matter and scalar field is given by 

\begin{equation}\label{47A}
    I_{total}=c_1\int R\sqrt{-g} d^4 x+\int L_m \sqrt{-g} d^4 x+\int L_s(\phi,\phi_{,\gamma}) \sqrt{-g}d^4 x,
\end{equation}
where $c_1$ is the numeric constant and $ \phi $ is the scalar field. The variational principle with respect to $\delta g_{\gamma\delta}$ gives

\begin{equation}\label{48A}
    R^{\gamma\delta}-\frac{1}{2}R g^{\gamma\delta}=\kappa (T^{\gamma\delta}_m+T^{\gamma\delta}_s),
\end{equation}
where $T^{\gamma\delta}_m=(\rho+p)u^\gamma u^\delta-p g^{\gamma\delta}$ is the matter energy momentum tensor and $T^{\gamma\delta}_s$ is the energy momentum tensor of scalar field. Then the matter action variation with respect to $\delta g_{\gamma\delta}$ is $ \delta I_m=\int T^{\gamma\delta}_m \sqrt{-g} \delta g_{\gamma\delta} d^4 x $ and the scalar field action variation with respect to $\delta g_{\gamma\delta}$ is $ \delta I_s=\int T^{\gamma\delta}_s \sqrt{-g} \delta g_{\gamma\delta} d^4 x $, where $T^{\gamma\delta}_s \sqrt{-g}=\frac{\partial(L_s \sqrt{-g})}{\partial g_{\gamma\delta}}-\frac{\partial}{\partial x^k} \frac{\partial(L_s \sqrt{-g})}{\partial g_{\gamma\delta , k}}$ is the scalar field energy momentum tensor. \\

Thus the Bianchi identity,
\begin{equation}\label{49A}
    (R^{\gamma \delta}-\frac{1}{2} R g^{\gamma\delta})_{; \delta} =0
\end{equation}
gives
\begin{equation}\label{50A}
    T^{\gamma \delta}_{m;\delta}=-T^{\gamma\delta}_{s;\delta}
\end{equation}

or equivalently
\begin{equation}\label{51A}
    \left[ (\rho+p)u^\delta\right]_{; \delta}u^\gamma +(\rho+p)u^\delta u^\gamma_{; \delta}-p^{'\gamma}=-T^{\gamma\delta}_{s;\delta},
\end{equation}

contracting both sides w.r.t. $ u_\gamma $ and using $ u_\gamma u^\gamma=1 $, $ u_\gamma u^\gamma_{;\delta}=0 $, we get the modified equation of continuity as:
\begin{equation}\label{52A}
   (\rho+p)u^\delta_{;\delta} -p_{,\delta}u^\delta =-T^{\gamma\delta}_{s;\delta} u_\gamma,
\end{equation}
or equivalently
\begin{equation}\label{53A}
   (\rho+p)u^\delta \sqrt{-g} =p_{,\delta}u^\delta \sqrt{-g}-T^{\gamma\delta}_{s;\delta} u_\gamma \sqrt{-g}.
\end{equation}
The appearance of the scalar field energy-momentum tensor of the right-hand side in the form of the term $ -T^{\gamma\delta}_{s;\delta}u_\gamma \sqrt{-g} $ shows that the presence of the scalar field does indeed affect the equation of continuity of the matter field. 

Therefore, the equation of continuity can be written as:
\begin{equation}\label{54A}
    \partial_\delta T^{\gamma \delta}_m +  T_m^{\alpha\delta} \Gamma^\gamma_{\alpha\delta} + T^{\gamma \alpha}_m\Gamma^\delta_{\alpha\delta}=-T^{\gamma\delta}_{s;\delta} u_\gamma,
\end{equation}
on expanding Eq. \ref{54A} we get the modified equation of continuity as:
\begin{equation} \label{55A}
   \dot{\rho}+3H(\rho+p) = -T^{\gamma\delta}_{s;\delta} u_\gamma \sqrt{-g}
\end{equation}



\begin{thebibliography}{99}

\bibitem{Canuto:1977zz}
V.~Canuto, S.~H.~Hsieh and P.~J.~Adams,
Phys. Rev. Lett. \textbf{39} (1977), 429-432.
	
\bibitem{Canuto:1977dp}
V.~Canuto, P.~J.~Adams, S.~H.~Hsieh and E.~Tsiang,
Phys. Rev. D \textbf{16} (1977), 1643-1663.
	
\bibitem {Wes} C. M. Wesson., Gravity Particles and Astrophysics (New York, Reidel, Dordrecht Holland, 1980)
	
\bibitem{Will:1984qgz}
C.~M.~Will,
Phys. Rept. \textbf{113} (1984), 345-422.
	
\bibitem{Singh:2013gwj}
J.~K.~Singh and N.~K.~Sharma,
Int. J. Theor. Phys. \textbf{53} (2014) no.2, 461-468.
	
\bibitem{Sharma:2014kzy}
N.~K.~Sharma and J.~K.~Singh,
Int. J. Theor. Phys. \textbf{53} (2014) no.12, 4132-4140.

\bibitem{Reddy:2007zzf}
D.~R.~K.~Reddy and R.~L.~Naidu,
Int. J. Theor. Phys. \textbf{46} (2007), 2788-2794.

\bibitem{Zeyauddin:2012bg}
M.~Zeyauddin and B.~Saha,
Astrophys. Space Sci. \textbf{343} (2013), 445-450.

\bibitem{Beesham:1998ih}
A.~Beesham,
Mod. Phys. Lett. A \textbf{13} (1998), 805-810.

\bibitem{Zeyauddin:2010zz} M.~Zeyauddin and S.~Ram,
Fizika B \textbf{19} (2010), 149-160.

\bibitem{Tawfik:2019dda}
A.~N.~Tawfik and E.~A.~El Dahab,
Grav. Cosmol. \textbf{25}, no.2, 103-115 (2019).

\bibitem{Chattopadhyay:2020mqj}
S.~Chattopadhyay, A.~Pasqua, A.~N.~Tawfik and R.~Myrzakulov,
Phys. Scripta \textbf{95}, no.8, 085005 (2020).

\bibitem{Tawfik:2017ngn}
A.~Tawfik and E.~Abou El Dahab,
Int. J. Theor. Phys. \textbf{56}, no.7, 2122-2139 (2017).

\bibitem{Tawfik:2011sh}
A.~Tawfik and T.~Harko,
Phys. Rev. D \textbf{85}, 084032 (2012).

\bibitem{Tawfik:2010bm}
A.~Tawfik, M.~Wahba, H.~Mansour and T.~Harko,
Annalen Phys. \textbf{523}, 194-207 (2011).

\bibitem{Tawfik:2010ht}
A.~Tawfik,
Can. J. Phys. \textbf{88}, 825-831 (2010).

\bibitem{Tawfik:2019jsa}
A.~N.~Tawfik and I.~Mishustin,
J. Phys. G \textbf{46}, no.12, 125201 (2019).

\bibitem{Tawfik:2011gh}
A.~Tawfik and H.~Magdy,
Can. J. Phys. \textbf{90}, 433-440 (2012).


\bibitem{Pradhan:2006rn}
A.~Pradhan and S.~Otarod,
Astrophys. Space Sci. \textbf{306} (2006), 11-16.
		
\bibitem{Pradhan:2012he}
A.~Pradhan,
Indian J. Phys. \textbf{88} (2014), 215-223.

\bibitem{Akarsu:2011zd}
O.~Akarsu and T.~Dereli,
Int. J. Theor. Phys. \textbf{51} (2012), 612-621.
	
\bibitem{Pradhan:2012zt}
A.~Pradhan,
Res. Astron. Astrophys. \textbf{13} (2013), 139-158.

\bibitem{Nagpal:2019vre}
R.~Nagpal, J.~K.~Singh, A.~Beesham and H.~Shabani,
Annals Phys. \textbf{405} (2019), 234-255.

\bibitem{Singh:2018xjv}
J.~K.~Singh, K.~Bamba, R.~Nagpal and S.~K.~J.~Pacif,
Phys. Rev. D \textbf{97} (2018) no.12, 123536.

\bibitem{Singh:2022jue}
J.~K.~Singh, H.~Balhara, K.~Bamba and J.~Jena,
JHEP \textbf{03} (2023), 191
[erratum: JHEP \textbf{04} (2023), 049].

\bibitem{Singh:2023gxd}
J.~K.~Singh, Shaily, A.~Singh, A.~Beesham and H.~Shabani,
Annals Phys. \textbf{455} (2023), 169382.

\bibitem{Singh:2024kez}
J.~K.~Singh, H.~Balhara, Shaily and P.~Singh,
Astron. Comput. \textbf{46} (2024), 100795.

\bibitem{Ellis:2013iea}
J.~Ellis, M.~Fairbairn and M.~Sueiro,
JCAP \textbf{02}, 044 (2014).

	
\bibitem{beesham} 
A.~Beesham, IJMMS \textbf{14}, no. 2, 305-308(1991).

\bibitem{Singh:2019fga}
K.~M.~Singh, S.~Mandal, L.~P.~Devi and P.~K.~Sahoo,
New Astron. \textbf{77}, 101353 (2020)


\bibitem{VargasdosSantos:2015kfv}
M.~Vargas dos Santos, R.~R.~R.~Reis and I.~Waga,
JCAP \textbf{02} (2016), 066.

\bibitem{SupernovaSearchTeam:1998fmf}
A.~G.~Riess \textit{et al.} [Supernova Search Team],
Astron. J. \textbf{116} (1998), 1009-1038.
	
\bibitem{SupernovaSearchTeam:2001qse}
A.~G.~Riess \textit{et al.} [Supernova Search Team],
Astrophys. J. \textbf{560} (2001), 49-71.
	
\bibitem{SupernovaCosmologyProject:1997zqe}
S.~Perlmutter \textit{et al.} [Supernova Cosmology Project],
Nature \textbf{391} (1998), 51-54.
	
\bibitem{SupernovaCosmologyProject:1998vns}
S.~Perlmutter \textit{et al.} [Supernova Cosmology Project],
Astrophys. J. \textbf{517} (1999), 565-586.
	
\bibitem{SupernovaCosmologyProject:2003dcn}
R.~A.~Knop \textit{et al.} [Supernova Cosmology Project],
strophys. J. \textbf{598} (2003), 102.
	
\bibitem{SupernovaSearchTeam:2003cyd}
J.~L.~Tonry \textit{et al.} [Supernova Search Team],
Astrophys. J. \textbf{594} (2003), 1-24.
	
\bibitem{HighZSNSearch:2005xhg}
A.~Clocchiatti \textit{et al.} [High Z SN Search],
Astrophys. J. \textbf{642} (2006), 1-21.
	
\bibitem{WMAP:2003ivt}
C.~L.~Bennett \textit{et al.} [WMAP],
Astrophys. J. Suppl. \textbf{148} (2003), 1-27.

\bibitem{Boomerang:2000efg}
P.~de Bernardis \textit{et al.} [Boomerang],
Nature \textbf{404} (2000), 955-959.
 
\bibitem{Hanany:2000qf}
S.~Hanany, P.~Ade, A.~Balbi, J.~Bock, J.~Borrill, A.~Boscaleri, P.~de Bernardis, P.~G.~Ferreira, V.~V.~Hristov and A.~H.~Jaffe, \textit{et al.}
Astrophys. J. Lett. \textbf{545} (2000), L5.
	
\bibitem{SupernovaSearchTeam:2004lze}
A.~G.~Riess \textit{et al.} [Supernova Search Team],
Astrophys. J. \textbf{607} (2004), 665-687.

\bibitem{Riess:2006fw}
A.~G.~Riess, L.~G.~Strolger, S.~Casertano, H.~C.~Ferguson, B.~Mobasher, B.~Gold, P.~J.~Challis, A.~V.~Filippenko, S.~Jha and W.~Li, \textit{et al.}
Astrophys. J. \textbf{659} (2007), 98-121.
	
\bibitem{SNLS:2005qlf}
P.~Astier \textit{et al.} [SNLS],
Astron. Astrophys. \textbf{447} (2006), 31-48.
	
\bibitem{Davis:2007na}
T.~M.~Davis, E.~Mortsell, J.~Sollerman, A.~C.~Becker, S.~Blondin, P.~Challis, A.~Clocchiatti, A.~V.~Filippenko, R.~J.~Foley and P.~M.~Garnavich, \textit{et al.}
Astrophys. J. \textbf{666} (2007), 716-725.
	
\bibitem{Amendola:2002kd}
L.~Amendola,
Mon. Not. Roy. Astron. Soc. \textbf{342} (2003), 221-226.

\bibitem{Blandford:2004ah}
R.~D.~Blandford, M.~A.~Amin, E.~A.~Baltz, K.~Mandel and P.~J.~Marshall,
ASP Conf. Ser. \textbf{339} (2005), 27.

\bibitem{Tawfik:2019qyd}
A.~N.~Tawfik and C.~Greiner,
Int. J. Mod. Phys. E \textbf{30}, no.08, 2150067 (2021).

\bibitem{Tawfik:2021rvv}
A.~N.~Tawfik and C.~Greiner,
Entropy \textbf{23}, no.3, 295 (2021)

\bibitem{Chawla:2012it}
C.~Chawla, R.~K.~Mishra and A.~Pradhan,
Eur. Phys. J. Plus \textbf{127} (2012), 137.

\bibitem{Mishra:2013lja}
R.~K.~Mishra, A.~Pradhan and C.~Chawla,
Int. J. Theor. Phys. \textbf{52} (2013), 2546-2559.

\bibitem{Ahmed:2013bdq}
N.~Ahmed and A.~Pradhan,
Int. J. Theor. Phys. \textbf{53} (2014), 289-306.

\bibitem{Tiwari:2017emt}
R.~K.~Tiwari, A.~Beesham and A.~Pradhan,
Grav. Cosmol. \textbf{23} (2017) no.4, 392-400.

\bibitem{Singh:2022nfm}
J.~K.~Singh, Shaily, S.~Ram, J.~R.~L.~Santos and J.~A.~S.~Fortunato,
Int. J. Mod. Phys. D, \textbf{32}, no. 7, 2350040 (2023).

\bibitem{Singh:2023ryd}
J.~K.~Singh, P.~Singh, E.~N.~Saridakis and S.~Myrzakul,
[arXiv:2304.03783 [gr-qc]].

\bibitem{Pan-STARRS1:2017jku}
D.~M.~Scolnic \textit{et al.} [Pan-STARRS1],
Astrophys. J. \textbf{859}, no.2, 101 (2018).

\bibitem{Riess:1998dv}
A.~G.~Riess, R.~P.~Kirshner, B.~P.~Schmidt, S.~Jha, P.~Challis, P.~M.~Garnavich, A.~A.~Esin, C.~Carpenter, R.~Grashius and R.~E.~Schild, \textit{et al.}
Astron. J. \textbf{117}, 707-724 (1999).

\bibitem{Jha:2005jg}
S.~Jha, R.~P.~Kirshner, P.~Challis, P.~M.~Garnavich, T.~Matheson, A.~M.~Soderberg, G.~J.~M.~Graves, M.~Hicken, J.~F.~Alves and H.~G.~Arce, \textit{et al.}
Astron. J. \textbf{131}, 527-554 (2006).

\bibitem{Hicken:2009df}
M.~Hicken, P.~Challis, S.~Jha, R.~P.~Kirsher, T.~Matheson, M.~Modjaz, A.~Rest and W.~M.~Wood-Vasey,
Astrophys. J. \textbf{700}, 331-357 (2009).

\bibitem{Contreras:2009nt}
C.~Contreras, M.~Hamuy, M.~M.~Phillips, G.~Folatelli, N.~B.~Suntzeff, S.~E.~Persson, M.~Stritzinger, L.~Boldt, S.~Gonzalez and W.~Krzeminski, \textit{et al.}
Astron. J. \textbf{139}, 519-539 (2010).

\bibitem{SDSS:2014irn}
M.~Sako \textit{et al.} [SDSS],
Publ. Astron. Soc. Pac. \textbf{130}, no.988, 064002 (2018).

\bibitem{Shaily:2024nmy}
Shaily, A.~Singh, J.~K.~Singh and S.~Ray,
[arXiv:2402.01780 [gr-qc]].

\bibitem{Balhara:2023mgj}
H.~Balhara, J.~K.~Singh and E.~N.~Saridakis,
[arXiv:2312.17277 [gr-qc]].

\bibitem{Balhara:2023owb}
H.~Balhara, J.~K.~Singh and J.~Jena,
[arXiv:2311.11926 [gr-qc]].

\bibitem{Singh:2022ptu}
J.~K.~Singh, Shaily, R.~Myrzakulov and H.~Balhara,
New Astron. \textbf{104} (2023), 102070.

\bibitem{Shaily:2024xho}
Shaily, A.~Singh, J.~K.~Singh and S.~Hussain,
[arXiv:2402.08709 [gr-qc]].


\bibitem{Gelman:1992zz}
A.~Gelman and D.~B.~Rubin,
Statist. Sci. \textbf{7} (1992), 457-472

\bibitem{Planck:2018vyg}
N.~Aghanim \textit{et al.} [Planck],
Astron. Astrophys. \textbf{641} (2020), A6
[erratum: Astron. Astrophys. \textbf{652} (2021), C4].

\bibitem{Santos:2005pe}
J.~Santos and J.~S.~Alcaniz,
Phys. Lett. B \textbf{619} (2005), 11-16.
	
\bibitem{Santos:2007zza}
J.~Santos, J.~S.~Alcaniz, M.~J.~Reboucas and N.~Pires,
Phys. Rev. D \textbf{76} (2007), 043519.
	
\bibitem{Sen:2007ep}
A.~A.~Sen and R.~J.~Scherrer,
Phys. Lett. B \textbf{659} (2008), 457-461.
	
\bibitem{Singh:2010zze}
J.~K.~Singh and N.~K.~Sharma,
Astrophys. Space Sci. \textbf{327} (2010), 293-298.
	
\bibitem{Singh:2010zzo}
J.~K.~Singh,
Mod. Phys. Lett. A \textbf{25} (2010), 2363-2371.

\bibitem{Mandal:2022ata}
S.~Mandal, P.~K.~Sahoo and J.~R.~L.~Santos,
Phys. Rev. D \textbf{106}, no.4, 048502 (2022).
	
\bibitem{Padmanabhan:2002vv}
T.~Padmanabhan and T.~R.~Choudhury,
Mon. Not. Roy. Astron. Soc. \textbf{344} (2003), 823-834.
	
\bibitem{carr} S. M. Carroll, \textbf{ Spacetime and Geometry: An Introduction to General Relativity}, Addison Wesley, 2004

\bibitem{Bolotin:2015dja}
Y.~L.~Bolotin, V.~A.~Cherkaskiy, O.~A.~Lemets, D.~A.~Yerokhin and L.~G.~Zazunov,
[arXiv:1502.00811 [gr-qc]] (2015).

\bibitem{Visser:1997qk}
M.~Visser,
Science \textbf{276} (1997), 88-90.

\bibitem{Visser:1997tq}
M.~Visser,
Phys. Rev. D \textbf{56} (1997), 7578-7587.

\bibitem{Singh:2022eun}
J.~K.~Singh, A.~Singh, G.~K.~Goswami and J.~Jena,
Annals Phys. \textbf{443} (2022), 168958.


\bibitem{Singh:2022wwa}
J.~K.~Singh, A.~Singh, Shaily and J.~Jena,
Chin. J. Phys. \textbf{86}, 616-627 (2023).
   
\bibitem{Sahni:2008xx}
V.~Sahni, A.~Shafieloo and A.~A.~Starobinsky,
Phys. Rev. D \textbf{78} (2008), 103502.
	
\bibitem{Sahni:2002fz}
V.~Sahni, T.~D.~Saini, A.~A.~Starobinsky and U.~Alam,
JETP Lett. \textbf{77} (2003), 201-206.
	
\bibitem{Alam:2003sc}
U.~Alam, V.~Sahni, T.~D.~Saini and A.~A.~Starobinsky,
Mon. Not. Roy. Astron. Soc. \textbf{344} (2003), 1057.
		
\bibitem{Hawking:1975vcx}
S.~W.~Hawking,
Commun. Math. Phys. \textbf{43} (1975), 199-220
[erratum: Commun. Math. Phys. \textbf{46} (1976), 206].

\bibitem{Singh:2019uwv}
C.~P.~Singh and S.~Kaur,
Phys. Rev. D \textbf{100}, no.8, 084057 (2019).

\bibitem{Brevik:2020psn}
I.~Brevik and A.~V.~Timoshkin,
Int. J. Mod. Phys. D \textbf{30}, no.02, 2150008 (2021).

\end{thebibliography}
\end{document}